\documentclass[a4paper,fleqn]{cas-sc}
\usepackage{tabularx}
\usepackage{longtable}
\usepackage[numbers, sort&compress]{natbib}
\usepackage[utf8]{inputenc}
\usepackage{xcolor}
\usepackage[colorinlistoftodos]{todonotes}
\begin{document}
\let\WriteBookmarks\relax
\def\floatpagepagefraction{1}
\def\textpagefraction{.001}

\shorttitle{Disaster Informatics after the COVID-19 Pandemic}

\shortauthors{Tran et~al.}

\title [mode = title]{Disaster Informatics after the COVID-19 Pandemic: Bibliometric and Topic Analysis based on Large-scale Academic Literature}  

\tnotemark[1]

\tnotetext[1]{This study has been partially supported by the Sarah Law Kennerly Endowed Funds and the Research Seed Grant from College of Information at University of North Texas.}

%\author[1,3]{CV Radhakrishnan}[type=editor, auid=000,bioid=1, prefix=Sir,role=Researcher,  orcid=0000-0001-7511-2910]
\author[1]{Ngan Tran}[%    auid=000,bioid=1,
                   %     prefix=Sir,
                   %     role=Researcher,
                        orcid=0000-0001-2345-6789]

%\cormark[1]


\ead{NganTran4@my.unt.edu}

%\ead[url]{www.cvr.cc, cvr@sayahna.org}

\credit{Methodology, Data curation, Formal analysis, Visualization, Validation, Software, Writing - Original draft}

\affiliation[1]{organization={Department of Information Science, University of North Texas},
    %addressline={1155 Union Cir}, 
    city={Denton},
    % citysep={}, % Uncomment if no comma needed between city and postcode
    postcode={76205}, 
    state={TX},
    country={United States}}

\affiliation[2]{organization={Department of Data Science, University of North Texas},
    %addressline={1155 Union Cir}, 
    city={Denton},
    % citysep={}, % Uncomment if no comma needed between city and postcode
    postcode={76205}, 
    state={TX},
    country={United States}}

%\author[2,4]{Han Theh Thanh}[style=chinese]
\author[2]{Haihua Chen}[%    auid=000,bioid=1,
                   %     prefix=Sir,
                   %     role=Researcher,
                        orcid=0000-0002-7088-9752]
\cormark[1]
\ead{Haihua.Chen@unt.edu}
\credit{Conceptualization, Methodology, Writing – original draft, review, Supervision}
%\affiliation[1]{organization={Department of Information Science, University of North Texas}}
%\author[2,3]{CV Rajagopal}[% role=Co-ordinator,    suffix=Jr,   ]
\author[1]{Ana Cleveland}[%    auid=000,bioid=1,
                   %     prefix=Sir,
                   %     role=Researcher,
                        orcid=0000-0001-9867-9418]
\cormark[2]

\ead{Ana.Cleveland@unt.edu}
\credit{Conceptualization, Methodology, Writing – original draft, Project administration, Funding acquisition}

\author[1]{Yuhan Zhou}[%    auid=000,bioid=1,
                   %     prefix=Sir,
                   %     role=Researcher,
                        orcid=0009-0006-3655-1527]
\ead{yuhanzhou@my.unt.edu}

\credit{Visualization, Writing - review \& editing}

\cortext[cor1]{Corresponding author}
\cortext[cor2]{Principal corresponding author}

% \fntext[fn1]{This is the first author footnote. but is common to third
% \fntext[fn2]{Another author footnote, this is a very long footnote and
%   it should be a really long footnote. But this footnote is not yet
%   sufficiently long enough to make two lines of footnote text.}

% \nonumnote{This note has no numbers. In this work we demonstrate $a_b$
%   the formation Y\_1 of a new type of polariton on the interface
%   between a cuprous oxide slab and a polystyrene micro-sphere placed

\begin{abstract}
This study presents a comprehensive bibliometric and topic analysis of the disaster informatics literature published between January 2020 to September 2022. Leveraging a large-scale corpus and advanced techniques such as pre-trained language models and generative AI, we identify the most active countries, institutions, authors, collaboration networks, emergent topics, patterns among the most significant topics, and shifts in research priorities spurred by the COVID-19 pandemic. Our findings highlight (1) countries that were most impacted by the COVID-19 pandemic were also among the most active, with each country having specific research interests, (2) countries and institutions within the same region or share a common language tend to collaborate, (3) top active authors tend to form close partnerships with one or two key partners, (4) authors typically specialized in one or two specific topics, while institutions had more diverse interests across several topics, and (5) the COVID-19 pandemic has influenced research priorities in disaster informatics, placing greater emphasis on public health. We further demonstrate that the field is converging on multidimensional resilience strategies and cross-sectoral data-sharing collaborations or projects, reflecting a heightened awareness of global vulnerability and interdependency. Collecting and quality assurance strategies, data analytic practices, LLM-based topic extraction and summarization approaches, and result visualization tools can be applied to comparable datasets or solve similar analytic problems. By mapping out the trends in disaster informatics, our analysis offers strategic insights for policymakers, practitioners, and scholars aiming to enhance disaster informatics capacities in an increasingly uncertain and complex risk landscape.

\end{abstract}

% \includegraphics{figs/grabs.pdf}

\begin{highlights}
    \item Bibliometrics of disaster informatics on a large-scale corpus with 5,496 papers. 
    \item Topic extraction using BERTopic and topic evaluation based on five metrics. 
    \item Evaluating GPT-3.5 and Llama 2 for topic summarization. 
    \item Fine-grained and topic-specific bibliometric analysis on 12 most significant topics. 
\end{highlights}

% Keywords

\maketitle

\section{Introduction}

Disaster informatics involves the design, development, and application of information technologies to generate, gather, process, store, and disseminate critical data that inform disaster preparedness, mitigation, response, and recovery efforts \citep{Ogie2020DisasterIA}. Within this multidisciplinary domain, it is essential to understand how research adapts in the face of global crises. The COVID-19 pandemic, for instance, has induced widespread disruptions across all facets of society, offering a unique opportunity to examine how disaster informatics research evolves, identifies new priorities, and responds to emerging challenges. Insights gained from this process can enhance the field’s capacity to mitigate the impacts of crises, facilitate recovery efforts, and strengthen resilience against future pandemics.

The COVID-19 pandemic triggered an unprecedented surge of publications in disaster informatics, as indicated by our preliminary search in PubMed Central using the keywords ``disaster informatics'', `` pandemic crisis'', ``crisis informatics''. This dramatic increase in literature suggests the development of new research themes and an intensified focus on public health crises within the field. Concurrently, the literature reflects evolving disaster management practices \citep{Potutan_Arakida_2021} and the introduction of a wide range of IT tools for emergency management during COVID-19 \citep{Asadzadeh_Ferdousi_2020}. However, critical gaps persist—particularly the limited integration between disaster informatics and public health frameworks \citep{Potutan_Arakida_2021}, as well as a lack of comprehensive pre-pandemic research on preparedness, mitigation, and prevention \citep{Asadzadeh_Ferdousi_2020}. Addressing these gaps is a key motivation for this study, which focuses on how the COVID-19 pandemic influenced research priorities in disaster informatics.

One major objective of this study is to map the landscape of disaster informatics research after the onset of COVID-19 through a large-scale bibliometric analysis. We aim to identify publication patterns, emerging topics, and key contributors—including authors, institutions, and countries. Although previous studies have conducted similar bibliometric analyses, they have often faced constraints: (1) reliance on a single database \citep{Ranjbari_et_al_2023, Ejaz_et_al_2022, Tan_Hao_2022, sahil2021bibliometric, Rana_2020, Sweileh_2019}; (2) lack of rigorous data quality assurance \citep{Rana_2020}, which is critical for bibliometric analyses \citep{nguyen2022assessing}; (3) dependence on co-occurring keywords for topic modeling \citep{chen2021demystifying}; and (4) inaccuracies in determining the most productive authors due to name ambiguity \citep{sahil2021bibliometric, Ejaz_et_al_2022}. We address these limitations by (1) sourcing data from multiple repositories (PubMed Central, Scopus, and Web of Science), ensuring a more comprehensive dataset; (2) using automated tools (Regex and Python) to efficiently detect and remove duplicate records and manually removing irrelevant data and adding missing data; (3) employing contextual topic modeling techniques that leverage titles and abstracts to generate more reliable and nuanced topic clusters; and (4) identifying authors by their full names to ensure accurate attribution of contributions.

Our second objective is to extract and summarize key research topics in disaster informatics from January 2020 to September 2022, leveraging large language models (LLMs) for topic modeling and summarization. LLMs, trained on extensive text corpora, excel in natural language understanding, text generation, and summarization tasks. While conventional topic modeling methods such as LDA or BERTopic effectively identify thematic structures, generative AI (GenAI) models—such as GPT—can produce human-readable summaries that clarify and condense complex findings. Although emerging approaches, including attention mechanisms \citep{dar2024contrastive}, NLP-driven analyses \citep{ma2024surveying}, and active learning strategies \citep{hanny2024active}, have been explored, their application in disaster informatics remains limited. In this study, we compare BERTopic with three two-layer BERTopic models through a rigorous evaluation on different metrics and select the best model for topic extraction. We then employ GenAI to generate coherent, human-readable summaries of our results, thus enhancing the interpretability and utility of topic modeling outputs.

Another objective of this study is to conduct a fine-grained, topic-specific bibliometric analysis. This approach illuminates how different authors, institutions, and countries contribute to various aspects of pandemic-related challenges, offering a more granular understanding of the intellectual landscape and collaboration patterns within the field. The insights gained can support the coordination of international efforts, ensure more equitable distribution of research expertise, and guide strategic investments to strengthen disaster informatics as a global resource against future crises.

With these objectives, we aim to answer the following four research questions based on the disaster informatics literature from January 2020 to September 2022:

\begin{itemize}
    \item RQ1: What are the patterns of publication and collaborations between countries, institutions, and authors?% in disaster informatics from January 2020 to September 2022?
    
    %\item What are the patterns of collaborations between authors, institutions, and countries in disaster informatics from 2020 to September 2022?
    \item RQ2: How to extract and evaluate topics from the literature?
    \item RQ3: What are the most significant topics regarding disaster informatics after the COVID-19 pandemic? 
    %investigated in disaster informatics from January 2020 to September 2022?
    \item RQ4: How are the countries, institutions, and authors collaborating with each other across the most significant topics? 
\end{itemize}

% The contributions of this study are as follows:
% \begin{itemize}
% % condense each bullet into 1 sentence.
%     \item The application and evaluation of topic models underscores the potential for refining and adapting existing topic models.
%     \item The proposed Topic Embedding Similarity metric enables a more accurate assessment of topic models. %is useful in assessing the similarities among topics resulting from topic modeling. This metric 
%     \item The application and evaluation of LLMs (GPT-3.5-turbo and Llama 2) in generating topic descriptions helps streamline the traditionally labor-intensive approach. 
%     %By identifying the level of comprehensiveness and inter-agreement, we can identify the best candidate for describing topics, thereby enhancing the utility and efficiency of topic modeling in research. %number of abstracts aligned with each topic description
%     \item Our content analysis revealed how the COVID-19 pandemic influenced productivity and research areas in disaster informatics and highlighted the research interests of the most active countries, institutions, and authors. 
%     %thereby, encouraging collaborations and assisting in coordinating international efforts to address the broad impacts of global crises like COVID-19. % focused on different aspects of the pandemic's impact
% \end{itemize}

The main contributions of this paper are as follows:
\begin{itemize}
% condense each bullet into 1 sentence.
    \item We conducted a comprehensive bibliometric analysis on large-scale disaster informatics literature for identifying the countries, institutions, authors and their collaborations.
    \item We proposed novel and effective methods for extracting, evaluating, and summarizing topics from the literature using advanced AI techniques. 
    %By identifying the level of comprehensiveness and inter-agreement, we can identify the best candidate for describing topics, thereby enhancing the utility and efficiency of topic modeling in research. %number of abstracts aligned with each topic description
    \item We performed fine-grained analysis of topics contributed and collaborated by the most active countries, institutions and authors. 
    %thereby, encouraging collaborations and assisting in coordinating international efforts to address the broad impacts of global crises like COVID-19. % focused on different aspects of the pandemic's impact
\end{itemize}

% Paper structure
The rest of this paper is organized as follows: Section \ref{sec:related_work} reviews literature. Section \ref{sec:data_method} describes the methodology. Section \ref{sec:results} presents the results. Section \ref{sec:discussion} further discusses the results. Lastly, Section \ref{sec:conclusion} concludes the paper and discusses the limitations and future work.

\section{Related Work}
\label{sec:related_work}

\subsection{Disaster Informatics}

%Disaster informatics involves the design, development, and application of information and technology solutions that enable the collection, processing, storage, and dissemination of vital information throughout the cycle of disaster preparedness, mitigation, response, and recovery \citep{Ogie2020DisasterIA}. 
The terms ``disaster informatics'' and ``crisis informatics'' are often used interchangeably \citep{Ogie2020DisasterIA}. Closely related to disaster informatics, crisis informatics focuses on the complex interplay of people, organizations, information, and technology before, during, and after disasters and crises \citep{Hagar_2014}. While there are subtle differences in emphasis, both domains are concerned with understanding the impacts of crises, reducing associated risks, and enhancing the effectiveness of management strategies. To ensure comprehensive coverage, we incorporated both terms into our search criteria.

The United Nations Office for Disaster Risk Reduction (UNDRR) defines disasters—whether natural or human-made—as events that significantly disrupt communities, causing substantial human, material, economic, and environmental losses \citep{assembly2016report}. The Sendai Framework for Disaster Risk Reduction 2015–2030 further broadens the scope of disaster risk reduction to include environmental, technological, and biological hazards, such as infectious diseases and pandemics.\footnote{\url{https://www.preventionweb.net/sendai-framework/sendai-framework-at-a-glance}} This expanded perspective underscores the link between disaster informatics and public health, reinforcing the importance of addressing outbreaks and pandemics. Reflecting this, we included the term ``pandemic crisis'' among our search queries.

Before the COVID-19 pandemic, the disaster informatics literature largely concentrated on leveraging social media for disaster detection \citep{Samuels2018TheSO, Wang2018UrbanCD}, preparation \citep{Grasso2016CodifiedHF, Martn2017LeveragingTT}, and response \citep{Martn2017LeveragingTT, Kar2019UsingSM}. Studies examined social media’s role as a platform for communication, information dissemination, and coordination during emergencies \citep{Reuter2018SocialMI, Reuter_Kaufhold_2017, Lee2017EnablersIC, Tan_Hao_2022, Malik2021PublicHA, Chon_Kim_2022}. While some research explored public behaviors \citep{Seo2019AmplifyingPA} and responses to disease outbreaks like MERS and Ebola \citep{Pieri2018MediaFA, Graham2017EmergingVD, Sambala2017AnticipationAR}, disaster informatics had seldom emphasized public health considerations within its framework.

The onset of COVID-19 sparked a surge in studies examining public health strategies in crisis management. These ranged from analyses of country-specific interventions \citep{Wang_Mao_2021, Kim_Park_2021, Assefa_Woldeyohannes_Damme_2022, Choi_Cho_Kim_Hur_2020, Houser_2022} to forward-looking approaches focusing on risk assessment and mitigation \citep{Kim_Shinde_Lone_Palem_Ghodake_2021, Torri_Sbrogio_Francia_Ferro_2020}. In Asia, for example, pandemic-related disaster response measures included the adoption of digitalized early warning systems, decentralized evacuation protocols, social distancing strategies, and remote psychological first aid \citep{Potutan_Arakida_2021}. These adaptations revealed a need for more integrated strategies, particularly those bridging the gap between disaster management and public health systems.

The pandemic also drove the diversification of IT tools employed in emergency management. Asadzadeh et al. classified these tools into categories spanning detection and diagnosis, treatment approaches, protection strategies, and overarching management objectives \cite{Asadzadeh_Ferdousi_2020}. Specific applications include bioinformatics tools for drug discovery, AI-enabled solutions for diagnosis and surveillance, telemedicine platforms for monitoring and forecasting patient trends, and mobile applications for contact tracing and public health communication. Despite these advancements, the study identified a notable pre-COVID-19 gap in research on IT-driven solutions for mitigation, prevention, and preparedness. Addressing this gap remains crucial for strengthening disaster informatics and enhancing resilience against future crises.

\subsection{Bibliometric Analysis for Disaster Informatics}

Bibliometric analysis offers a quantitative framework for examining publications, enabling researchers to gain an overview of the literature, identify research trends, analyze publication patterns, and assess overall field performance \citep{Ellegaard_Wallin_2015}. By examining key contributors—such as authors, institutions, and countries—and uncovering the factors that shape publication trends and productivity, bibliometric methods clarify how knowledge evolves across different domains \citep{Ejaz_et_al_2022}. These approaches frequently incorporate citation analysis to trace the flow of information and link findings across studies, while content analysis techniques, including keyword extraction and topic modeling, reveal research hotspots and the thematic evolution of the field over time.

Several bibliometric studies have underscored the growing prominence of specific research trends. For example, Ranjbari et al. identified research gaps in post-COVID-19 waste management and proposed new directions for future inquiry \cite{Ranjbari_et_al_2023}. Similarly, Milán-García et al. compared climate change-induced migration studies across different databases, pinpointing key thematic areas \cite{MilnGarca2021ClimateCM}. In a longitudinal perspective, Irfan Ahmad Rana examined disaster and climate change resilience literature from 1992 to 2019, offering insights into temporal shifts, emerging research directions, and leading contributors in the field \cite{Rana_2020}.

At the same time, these studies highlight methodological limitations that can affect the reliability of bibliometric findings. For instance, Ejaz et al. analyzed the growth of Omicron-related publications, focusing on prolific authors, institutions, and journals \cite{Ejaz_et_al_2022}. Their study noted issues like incomplete text processing and inaccurate author identification, which could skew results. Similarly, Sahil and Sandeep Kumar Soo's examination of ICT applications in disaster management encountered difficulties in author identification, relying heavily on initials and potentially distorting collaboration network analyses \cite{sahil2021bibliometric}.

A critical gap emerging from existing bibliometric research involves the underrepresentation of public health dimensions in disaster informatics. Tan and Hao studied information behaviors during natural disasters but overlooked public health emergencies, even though the COVID-19 outbreak fell within the examined time frame \cite{Tan_Hao_2022}. This oversight highlights the persistent lack of integration between public health and disaster research. In fact, Sweileh noted that, before COVID-19, only 16\% of publications in the disaster informatics domain addressed health-related issues \cite{Sweileh_2019}. The current pandemic underscores the urgency of bridging this gap, demonstrating that disaster informatics must integrate public health considerations to enhance preparedness, response, and recovery in the face of global crises.

\subsection{Topic Modeling}

Topic modeling techniques facilitate the discovery, interpretation, and annotation of thematic structures within large document collections \citep{zhang2014review}, thereby enhancing our understanding of relationships among texts. Traditional methods include Latent Semantic Analysis (LSA), Non-Negative Matrix Factorization (NMF), and Latent Dirichlet Allocation (LDA) \citep{Yu2022TM}. LSA employs singular value decomposition (SVD) to reduce the dimensionality of the document-term matrix, revealing latent structures. However, its reliance on linear algebra limits its ability to capture non-linear relationships and higher-level semantics. LDA, by contrast, models each document as a probabilistic mixture of topics and each topic as a distribution of words \citep{blei2003latent}. Although LDA is widely adopted, determining the optimal number of topics remains challenging, and there is no universally accepted selection method \citep{gan2021selection, zhang2014review}. More broadly, traditional approaches struggle with capturing deeper semantic relationships, optimizing modeling performance, and scaling effectively to large datasets \citep{grootendorst2022bertopic}.

Recent advances in topic modeling have introduced methods that address these limitations. Top2Vec, for instance, employs Doc2Vec to produce document embeddings that more accurately reflect complex content relationships, allowing it to infer the number of topics automatically \citep{angelov2020top2vec}. BERTopic goes a step further, leveraging transformer-based embeddings to capture rich contextual information and enabling dynamic topic tracking over time \citep{electronics12122605}. This approach generates document embeddings using pre-trained language models, clusters documents based on embedding similarity, and identifies topic terms using a class-based TF-IDF procedure \citep{grootendorst2020keybert}. After reducing embedding dimensionality for efficient clustering, BERTopic uses the density-based algorithm HDBSCAN to form clusters without predefining their number. It has been successfully applied to social media content analysis \citep{hristova2022media, lindelof2023dynamics}, bibliometric studies \citep{raman2024unveiling}, and investigations of topic evolution \citep{wang2023identifying}.

Despite the growing popularity of BERTopic, comprehensive evaluations of its performance are still rare. Existing evaluation efforts rely on metrics such as topic coherence \citep{stevens2012exploring}, topic stability \citep{hosseiny2023review}, topic diversity \citep{dieng2020topic}, and classification accuracy \citep{terragni2020constrained}. Few studies have proposed a robust evaluation framework with validated metrics \citep{abdelrazek2023topic}. In the context of disaster informatics, Deng at al. applied topic modeling to identify and track public information needs during disasters, but their evaluation relied on manual topic interpretation and adjustments of topic granularity \cite{deng2020detecting}.

To advance the evaluation of topic modeling in disaster informatics, our study introduces the Topic Embedding Similarity metric, which quantifies similarities between topics using their embedded vector representations. We integrate this metric with four additional evaluation measures, providing a more comprehensive and reliable assessment of topic modeling results.

As previous research shows that public health was an underexplored area within disaster informatics before the COVID-19 pandemic, our study aims to fill this gap by combining bibliometric analysis, topic modeling, and generative AI (GenAI)-based content analysis. The following section details our research design and methodology.

\section{Methodology} 
\label{sec:data_method}

\subsection{Research Design}

\begin{figure}
    \centering
    \includegraphics[width=0.6\linewidth]{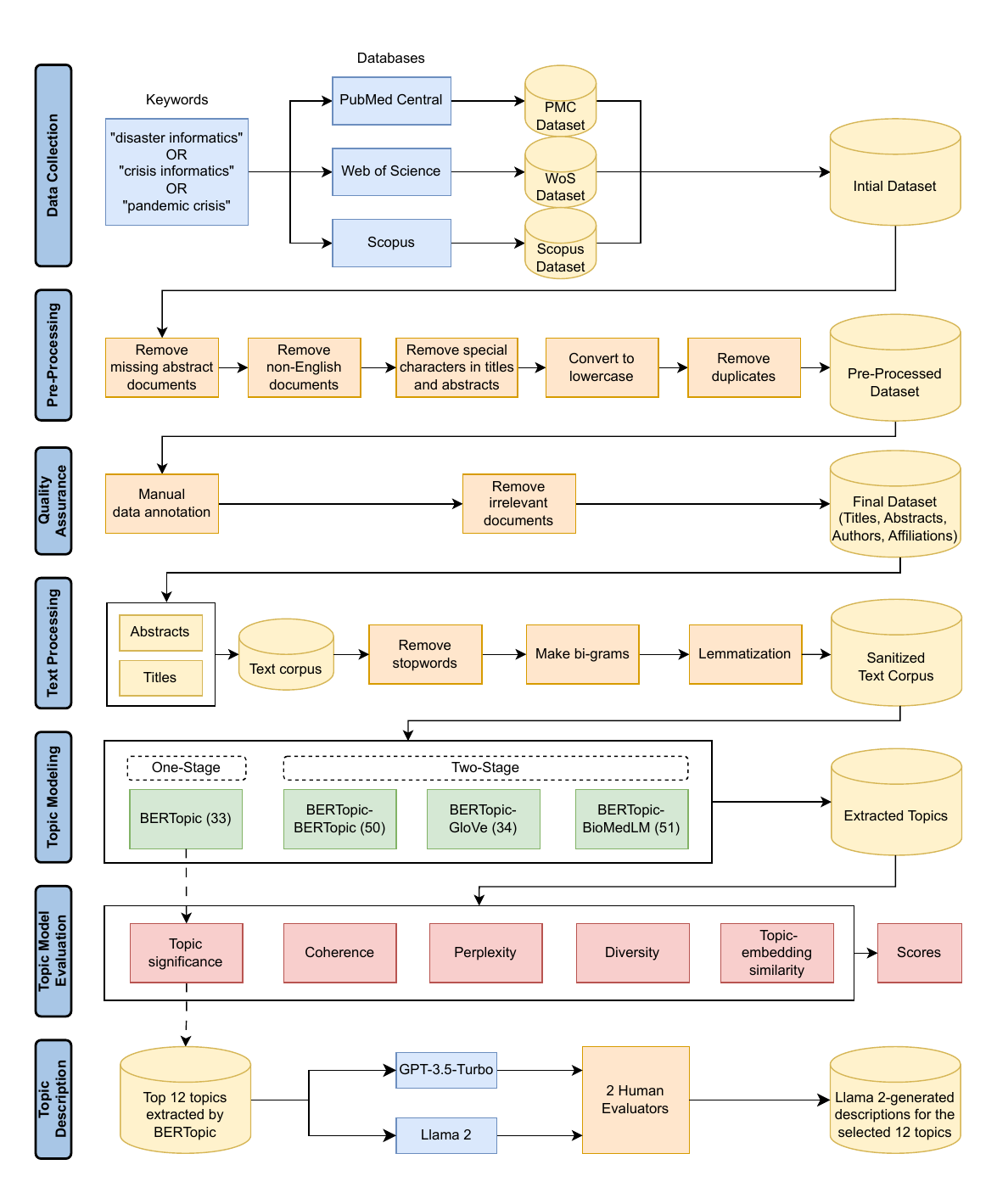} %ResearchDesign.png
    \caption{Research design.}
    \label{fig:ResearchDesign}
\end{figure}

Our research design, illustrated in Figure \ref{fig:ResearchDesign}, consists of seven phases: (1) data collection, (2) data pre-processing, (3) data quality assurance, (4) text processing, (5) topic modeling, (6) topic evaluation, and (7) topic description. Each phase is described in detail below.

\begin{enumerate}
    \item Data collection: We retrieved literature on disaster informatics published between January 2020 and September 2022 from three major academic databases: PubMed Central, Web of Science (WoS), and Scopus.
    \item Data pre-processing: We removed non-English documents, documents missing abstracts, and duplicate records. 
    \item Data quality assurance: To remove irrelevant data and ensure other information is correct, we invited master students in health informatics to annotate every document. A detailed guideline was developed for the annotation. This process ensured that our final dataset consisted of high-quality entries containing complete titles, abstracts, and authorship information.
    \item Text processing: The titles and abstracts of the curated dataset were concatenated to form a corpus of textual data suitable for topic modeling analysis.
    \item Topic modeling: We applied and compared four different BERT-based topic modeling methods to uncover the underlying thematic structures within the corpus. This step aimed to identify the most representative and meaningful topics across the collected literature.
    \item Topic evaluation: We quantitatively evaluated the topic models using five metrics: coherence, perplexity, diversity, topic-embedding similarity, and average topic significance. These metrics provided a comprehensive assessment of the quality, stability, and interpretability of the generated topics.
    \item Topic description: We employed and compared GPT-3.5-turbo and Llama 2 for generating summaries for the most significant topics. 
\end{enumerate}

\subsection{Data Collection and Pre-processing}
\begin{figure}
    \centering
    \includegraphics[width=0.6\linewidth]{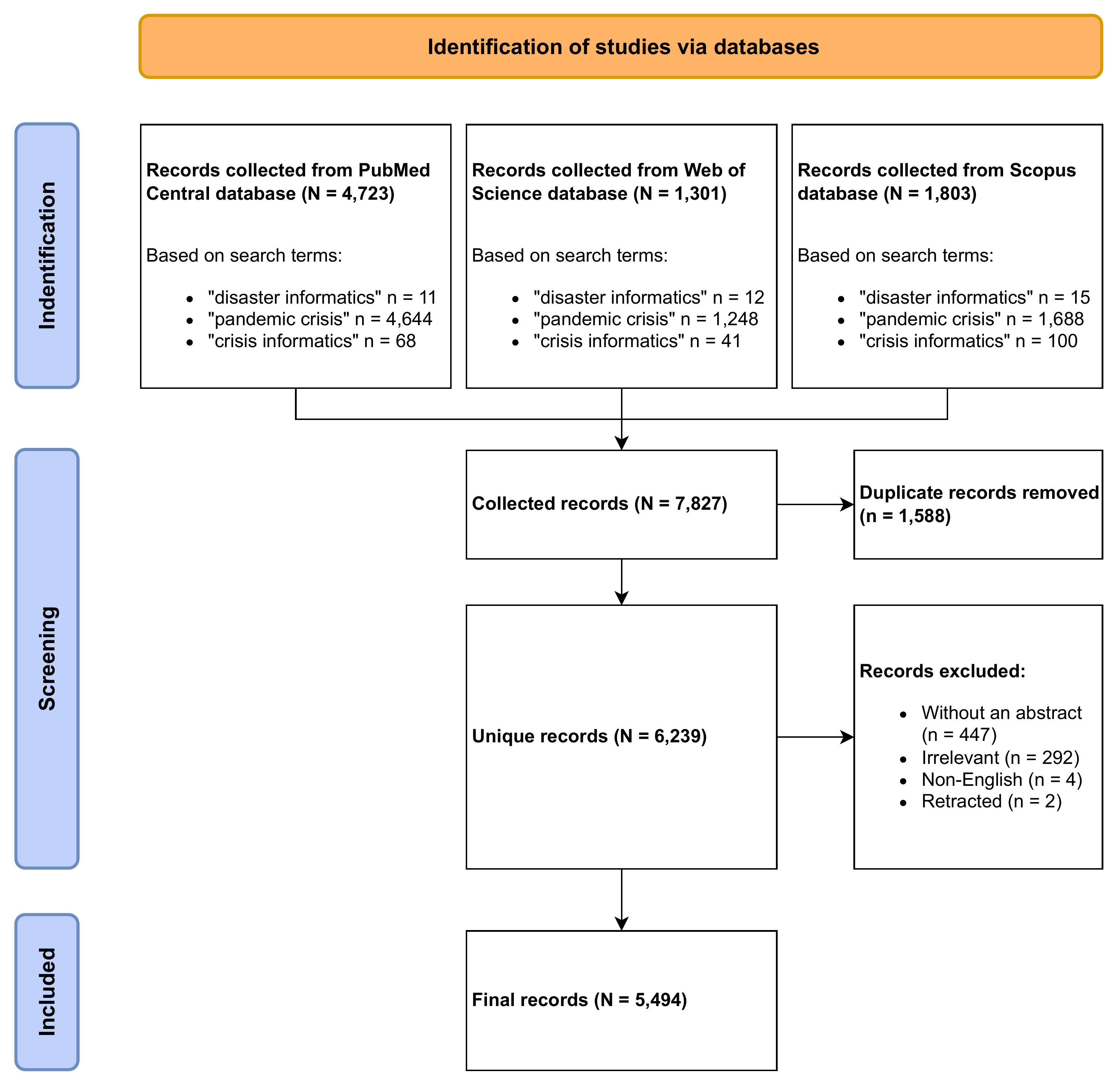} % PRISMA-V2.png
    \caption{PRISMA flow diagram (Data collected from January 2020 to September 2022).}
    \label{fig:prisma}
\end{figure}

% fix quotation mark.
We collected the literature from PubMed Central, Scopus, and WoS from January 2020 to September 2022 using three keywords: ``disaster informatics'', `` pandemic crisis'', and ``crisis informatics''. The collected documents are sometimes referred to as the post-pandemic literature. As shown in Figure \ref{fig:prisma}, we collected 7,827 papers in total for screening. For data pre-processing, we remove 1,588 duplicates, 447 without an abstract, 292 irrelevant, four non-English, and two retracted papers, resulting in a final dataset with 5,496 papers for bibliometrics and topic analysis.  

% \subsection{Data Pre-processing}

% %To ensure data quality, we identified and removed these duplicates, irrelevant documents, and those lacking abstracts in the next phase. 
% In the data pre-processing phase, we removed duplicates, documents without an abstract, and non-English documents. Before removing duplicates, we removed special characters from the titles and converted them to lowercase. This step allows us to identify and remove duplicates. There were 1,588 duplicates from the collected data across three databases. 

% Additionally, we removed 447 documents without an abstract as there is insufficient information for conducting topic modeling among documents without an abstract.

% Lastly, since the topic models used in this study were pre-trained with English content, including abstracts written in foreign languages may affect the topic modeling results. Hence, we identified 4 non-English abstracts and removed them.

\subsection{Data Quality Assurance}
The ``garbage in, garbage out'' principle suggests that data quality is the foundation for downstream tasks such as data analysis and machine learning \citep{rogers-2021-changing, zhou2024survey}. However, under-valuing of data work is common to all of data analysis and AI development \citep{sambasivan2021everyone}. In this paper, we implement data quality assurance to ensure that all documents are relevant to the scope of this study, which are relevant to disaster informatics, crisis informatics, and pandemic crisis. In the raw data, many publications contain formatting issues such as extra spacing, irrelevant details, and incomplete information in the affiliation fields, it complicates automatic data parsing. We addressed this by employing manual annotations from 6 human annotators to extract the names of institutions and countries in the affiliation field. We provided annotation training with written instructions and feedback and performed spot-checking to minimize annotation errors. In addition, we inspect each document: if a document does not contain one of terms like ``disaster'',  ``crisis'', ``pandemic'', or ``COVID-19'', we also consider it irrelevant and remove it from the dataset.  As a result, we removed 292 irrelevant along with 2 retracted publications. The result of this phase is the final dataset of 5,494 documents, with 1,016, 2,425, and 2,053 documents from 2020, 2021, 2022, respectively. % add number of publications per year here.
% add references to support why we need to ensure data quality assurance when using ML.  

\subsection{Text Processing}
\label{subsec-text-processing}
Text processing aims to prepare the input text for topic modeling. We combined the titles and abstracts from the final dataset to create a text corpus, and we removed stop words, created bi-grams, and conducted lemmatization. This process is demonstrated in Figure \ref{fig:ResearchDesign}. The output of this process is referred to as a sanitized text corpus.

\begin{itemize}
    \item Step 1: remove stop words that do not provide useful information in a document. We created a list of stop words by combining the stop words from the Gensim library and NLTK library in Python, and we added other redundant words that frequently occur in an abstract such as ``method'', ``background'', ``result'', among others.
    \item Step 2: create bi-grams, which are pairs of words that frequently occur together, and it is treated as a single word through concatenation. Bi-grams provide extra contextual information to a word, allowing us to generate meaningful insights from topic modeling results.
    \item Step 3: normalize words into their original form.  For example, the words ``running'' and ``ran'' become ``run'' after lemmatization. This technique reduces noises from the variations of a word, enabling us to identify keywords. Each abstract is lemmatized using the WordNetLemmatizer from the NLTK library in Python.
\end{itemize}

Based on the clean data, we describe the bibliometric analysis, topic modeling, and topic-wise bibliometric analysis strategies in the next three subsections. 

\subsection{Overall Bibliometric Analysis}
\label{subsec:overall-bib-analysis}

For publication output analysis, we counted unique documents per country, institution, and author and used the Python Matplotlib library to visualize them. A country is only counted once per document regardless of the number of authors affiliated with that country. Similarly, an institution is counted only once per article, irrespective of the number of authors affiliated with that institution. For authors affiliated with multiple institutions, only the first listed institution on the publication is considered the contributing institution. An author was referred to by their last name and first name. 

For collaboration analysis, we used Gephi to visualize the networks, in which a node is represented as an entity (country, institution, or author), and the links between two nodes represent the collaborations of two entities. The size of a node denotes the publication output of that particular entity, while the thickness of a link indicates the frequency of the collaborations between two entities. The clusters in Figures \ref{fig:countries-nw22}, \ref{fig:institutions-nw22}, and \ref{fig:authors-nw22} were identified using the Modularity function as a community detection tool in Gephi \citep{ICWSM09154}. To enhance the clarity of the networks, we excluded nodes with less than a publication threshold, which varies based on a specific network.

\subsection{Topic Extraction and Evaluation}
%Figure \ref{fig:topic_modeling} demonstrates our workflow in topic modeling, evaluation, and summarization. 
A topic is a specific area of interest that a collection of documents explores and reports knowledge. The sanitized text corpus was used as the input for topic modeling. We used four topic models: BERTopic, BERTopic-BERTopic, BERTopic-GloVe, and BERTopic-BioMedLM to extract topics from the collected literature.

To quantitatively evaluate the topics and identify the best-performing model, we used five metrics: coherence, perplexity, diversity, topic-embedding similarity, and topic significance (Section \ref{subsec:topic-modeling-eval}). Among the topics generated by the best-performing model, we selected the ones with the highest significant scores to qualitatively evaluate their levels of comprehensiveness. %(\ref{subsec:topic-description-eval}).

% In the next subsections, we provide more details regarding the structures of the selected topic models and our topic modeling strategy. 

\subsubsection{Topic Modeling Algorithms}
\label{subsec:topic-model-structure}

This study implemented and compared four topic models: BERTopic \citep{grootendorst2022bertopic}, BERTopic-BERTopic, BERTopic-GloVe, and BERTopic-BioMedLM. These models were established based on three main structures: % (Figure\ref{fig:modelstructure}): 
BERTopic, GloVe-HDBSCAN, and BioMedLM-HDBSCAN. 

BERTopic is a pre-trained topic modeling method that leverages the Transformer architecture for sentence embedding, UMAP for dimensionality reduction, HDBSCAN for clustering similar documents together, C-TF-IDF for topic representation \citep{grootendorst2022bertopic}, and KeyBERT \citep{grootendorst2020keybert} for keyword extraction. BERTopic was selected because it does not require users to specify the optimal number of topics, but requires an input of the minimum cluster size, which is 10 documents by default. 

%Due to the high volume of our data, we select a minimum cluster size, which is further described in section \ref{subsec-topic-modeling-strategies}. 
%BERTopic is a complete framework that takes in a list of documents and outputs the topics and their corresponding keywords. 
%Unlike other topic modeling methods that categorize documents into a predetermined number of clusters, BERTopic groups documents with similar content into clusters, defining any cluster meeting the minimum size requirement as a topic. The clusters with less than the minimum size and documents that do not belong to a cluster are considered outliers (group ``-1''). 

%unlike Latent Dirichlet Allocation (LDA) topic model (*cite LDA), thereby overcoming LDA weakness \citep{zhang2014review} 

% just mention the difference between the proposed models and BERTopic
We proposed the other two topic modeling structures, which used GloVe-HDBSCAN and BioMedLM-HDBSCAN as an embedding method. The GloVe-HDBSCAN model uses GloVe \citep{Pennington2014GloVeGV} for word embedding. The BioMedLM-HDBSCAN model uses BioMedLM for word embedding. BioMedLM is a language model created from GPT-2, previously known as PubMedGPT \citep{Bolton2024BioMedLMA2}. GloVe-HDBSCAN and BioMedLM-HDBSCAN have the same structure, and they both use a standard scaler to normalize the vector representation of each document, UMAP \footnote{\url{https://umap-learn.readthedocs.io/en/latest/}} for dimensionality reduction, HDBSCAN \footnote{\url{https://hdbscan.readthedocs.io/en/latest/index.html}} for clustering similar documents together, and KeyBERT \footnote{\url{https://github.com/MaartenGr/KeyBERT}} for keyword extraction. GloVe-HDBSCAN and BioMedLM-HDBSCAN function similarly to BERTopic. Both models take in a list of documents and the minimum cluster size and output the topics along with the corresponding keywords. 

%It uses an aggregated global word-word co-occurrence matrix, which was pre-trained on the GloVe-Wiki-Gigaword-300 embeddings, to obtain the vector representation of each word . BioMedLM was trained on 16 million abstracts and 5 million full-text articles from PubMed and PubMed Central databases in the Pile dataset. Besides the difference in the embedding methods,

We used GPT-3.5-turbo \footnote{\url{https://platform.openai.com/docs/models/gpt-3-5-turbo}} and Llama 2 \footnote{\url{https://www.llama.com/llama2/}} to generate a topic description based on a list of titles and abstracts in each topic. This step is described in Section \ref{subsec:topic-description}. 

%Upon analyzing our preliminary results, we decided not to use the keyword extraction tool KeyBERT for topic descriptions. This is because the extracted keywords do not provide sufficient contextual information for a comprehensive description. Instead,

\subsubsection{Topic Modeling Strategies}
\label{subsec-topic-modeling-strategies}
We employed two strategies for topic modeling. The first strategy conducts topic modeling only once (one-stage topic modeling), and we used BERTopic. The second strategy conducts topic modeling twice (two-stage topic modeling): BERTopic-BERTopic, BERTopic-GloVe, and BERTopic-BioMedLM. 

In one-stage modeling, BERTopic \citep{grootendorst2022bertopic} was used once to extract the topics with the minimum cluster size m. The default minimum cluster size is 10. However, given the size of our data (5,496 documents), we set m to be at least 30 documents. 

In two-stage modeling, we used topic modeling twice to identify the topics. In the first stage, we used BERTopic 
with a minimum cluster size of 30 documents (minimum cluster n=30) to group similar documents together. From the initial clusters, we applied the following topic models: BERTopic, GloVe-HDBSCAN, and BioMedLM-HDBSCAN, to extract the topics within each cluster. In the second stage, each topic must contain at least 15 documents (minimum cluster m=15). 
When BERTopic is used in the second stage, we refer to the method as BERTopic-BERTopic. Similarly, when GloVe-HDBSCAN or BioMedLM-HDBSCAN is used in the second stage, the method is referred to as BERTopic-GloVe or BERTopic-BioMedLM respectively.

\subsubsection{Topic Modeling Evaluation}
\label{subsec:topic-modeling-eval}

We evaluate the performance of the four topic modeling methods using five metrics: coherence, perplexity, diversity, topic-embedding similarity, and average topic significance.

\begin{itemize}
    \item \textbf{Coherence} assesses how interpretable a topic is. It determines the quality of the topics by evaluating whether the keywords of each topic are semantically related. Topic coherence quantifies the lexical similarity among pairs of top keywords. A logically consistent topic has a high coherence score. We calculate coherence using the top 10 keywords of each topic through the use of Gensim library in Python \footnote{\url{https://radimrehurek.com/gensim/models/coherencemodel.html}}. The overall coherence of a model is the average coherence across all topics.
    %However, the number of topics chosen can influence coherence, and there is a trade-off where optimizing for lower perplexity may lead to reduced coherence.
    
    \item \textbf{Perplexity} measures the model’s capability to cluster new documents based on the learned topics. A low perplexity indicates that the model is confident in its prediction, and vice versa. Perplexity is expressed in equation (\ref{eq:perplexity}) \citep{blei2003latent}, in which N indicates the number of words in the corpus, P(topic|doc) indicates the likelihood of a word document d, representing the document index ranging from 1 to M documents.
    \begin{equation}
        \label{eq:perplexity}
        \text{Perplexity} = \exp \left(-\frac{1}{N} \sum_{d=1}^M \log(P(\text{\textbf{w}}_{d}))\right) 
    \end{equation}
    
    \item \textbf{Topic Diversity} measures how different and unrelated the topics are from each other, which is necessary to comprehensively represent the corpus. Topic diversity is an important criterion for judging the quality of a learned model. In this paper, topic diversity is assessed through the percentage of unique words among the top 10 keywords across all topics (equation \ref{eq:topic_diversity}). The more diverse (a higher topic diversity score), i.e. dissimilar, the resulting topics are, the higher will be the coverage regarding the various aspects about disaster informatics. 
    \begin{equation}
        \label{eq:topic_diversity}
         \text{Topic Diversity} = \frac{\text{number of unique words}}{\text{number of keywords across all topics}}
    \end{equation}
    % Topic diversity can be affected by the pre-set number of topics. Too many topics can result in redundancy with overlapping themes, whereas too few can result in overly broad topics lacking specific details. 
    
    \item \textbf{Topic Embedding Similarity} was proposed to assess topic diversity through the similarity between the contextual representations of topics, rather than just their lexical characteristics. 
    Specifically, cosine similarity was calculated between the vector representations of each topic pair, as defined in equation (\ref{eq:cosine_similarity}), where A and B are the vector representations of two topics. A lower topic embedding similarity indicates that the topics contain unique concepts. The overall topic embedding similarity of a model is the average cosine similarity across all topic pairs. 
    \begin{equation}
        \label{eq:cosine_similarity}
        \text{Cosine Similarity}(\mathbf{A}, \mathbf{B}) = \frac{\mathbf{A} \cdot \mathbf{B}}{\|\mathbf{A}\| \|\mathbf{B}\|}
    \end{equation}

    \item \textbf{Topic Significance} assesses the significance of a topic through its distance from a uniform distribution of words \citep{topicsignificance}, and we used the implementation from OCTIS\footnote{\url{https://github.com/MIND-Lab/OCTIS}} \citep{terragni2020octis}.
    A uniform distribution represents a junk topic that contains many important terms but fails to provide specific information \citep{topicsignificance}. In contrast, a genuine topic tends to be skewed towards a smaller set of terms \citep{topicsignificance}, representing informative content. 
    Given that our collected literature covered various aspects of the pandemic's impact and included many terms from different fields, the topic significance metric is ideal for assessing topic models based on the information in the generated topics. A high topic significance score indicates a highly informative topic. The overall topic significance of a model is the average topic significance across all topics.
\end{itemize}

\subsubsection{Topic Description}
\label{subsec:topic-description}
Once we identified the best-performing topic model, we selected 12 topics with the highest topic significance scores for summarization and further analysis. These topics were chosen because they represent the most informative research areas with a strong thematic focus, making them ideal for in-depth analysis and interpretation. It enables us to identify specific research areas within the field of disaster informatics during the pandemic, and it allows us to explore insights into the effect of the pandemic. 

To get a description for each topic, we used GPT-3.5-turbo and Llama 2 to generate a summary based on a list of abstracts within the topic. Due to the excessive length of the abstracts within a topic, we divided them into several chunks and asked each GenAI model to summarize each chunk. Based on these summaries, each GenAI model gave a final summary, which was referred to as a topic description. To identify an effective GenAI model for this task, we asked two human experts to manually evaluate the generated topic descriptions. This evaluation helps us verify whether the abstracts of each topic share a common ground, aiding our understanding of the content of these 12 topics. %Ultimately, this strategy leverages both GenAI and human analysis, which requires less labor than manually crafting a description for each topic. 

%This helps us identify the most effective GenAI model for describing our topics, and requires  without extensive manual labor. 

%The generated topic descriptions were independently evaluated by two human evaluators. Each evaluator read through each abstract and assessed its alignment with the corresponding topic description by marking it as ``Yes'', ``No'', or ``Maybe'' and providing their rationales. Abstracts marked as ``Maybe'' were discussed further to reach a consensus on the final answer. These evaluations were encoded, with ``Yes'' translating to 1 and ``No'' to 0.

We assessed the effectiveness of each GenAI model in generating topic descriptions by calculating the level of comprehensiveness for each topic. This measure is defined as the proportion of abstracts that both evaluators unanimously rated as ``Yes'', relative to the total number of abstracts for each topic. The evaluation result of the generated descriptions is discussed in Section \ref{sec:topic-description-evaluation}. The GenAI with the highest average level of comprehensiveness across the 12 topics was considered the most effective one.  

\subsection{Topic-Wise Bibliometric Analysis}
To conduct bibliometric analysis for each of the top 12 topics, we first separated the data of each topic and then applied the same steps as described in Section \ref{subsec:overall-bib-analysis} to the data of each topic. 

\section{Results} %Bibliometric Analysis Result
\label{sec:results}

\subsection{Overall Bibliometric Analysis Results}
% define active authors.

%Figure \ref{fig:doc_freq} illustrates the yearly number of publications from January 2020 to September 2022. Each bar in the figure represents the number of publications each year, with the final bar showing the data only up to September 2022. From 2020 to 2021, the number of publications doubled from 1016 to 2425, and in 2022, the number of publications was 2053.
Out of 5,494 publications that were retrieved from January 2020 to September 2022, there are 1,016 publications in 2020, 2,425 in 2021, and 2,053 published until September 2022. The number of publications doubled from 2020 to 2021. 
A recent literature search showed that from September to December 2022, there were 2,920 additional publications. This indicates that the literature on disaster informatics keeps increasing afterward.

In the section below, we identify the most active countries, institutions, and authors contributing to the disaster informatics literature from January 2020 to September 2022, and analyze their collaborations. For the purpose of this study, the term ``active'' means those that produce the highest number of publications.  

\subsubsection{Countries}

\begin{figure}
    \centering
    \includegraphics[width=0.75\linewidth]{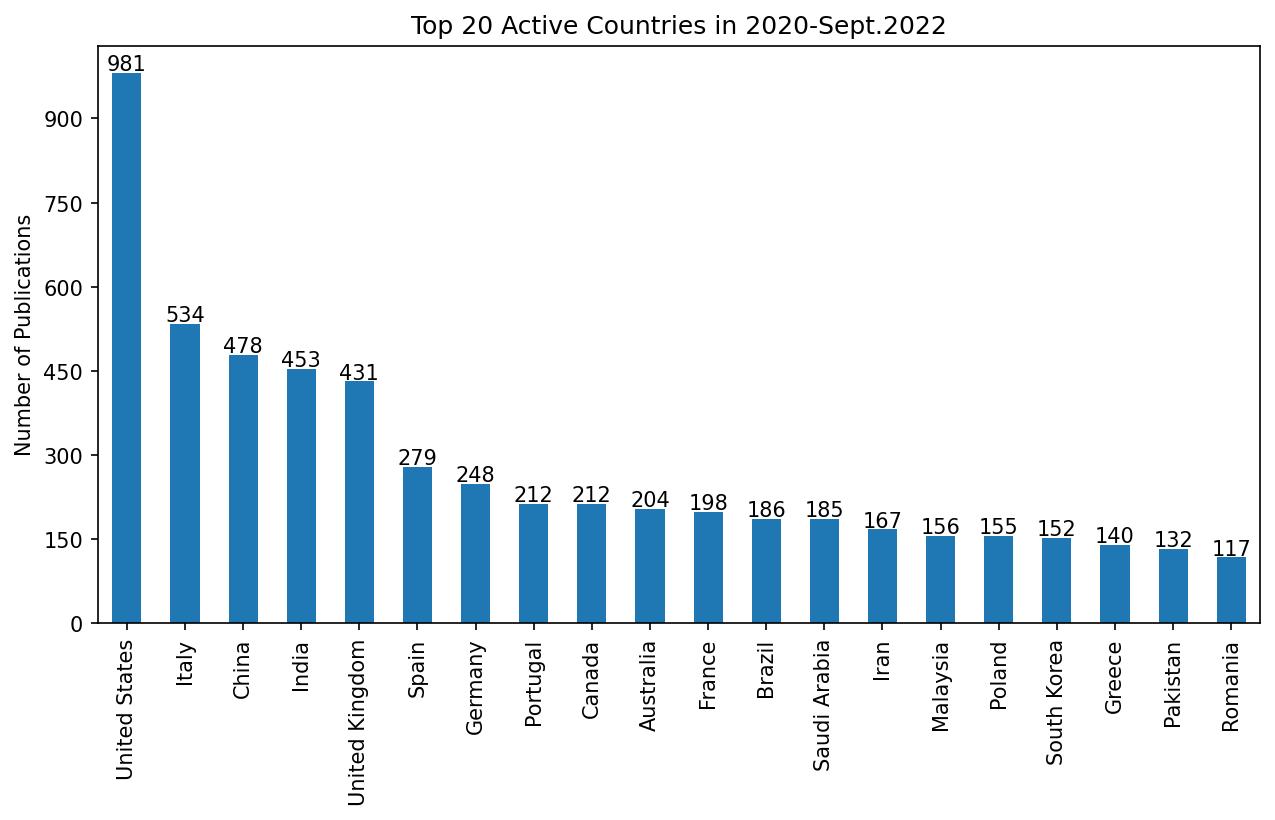}
    \caption{The most active countries.}
    \label{fig:top-countries22}
\end{figure}

\begin{figure}
    \centering
    \includegraphics[width=0.75\linewidth]{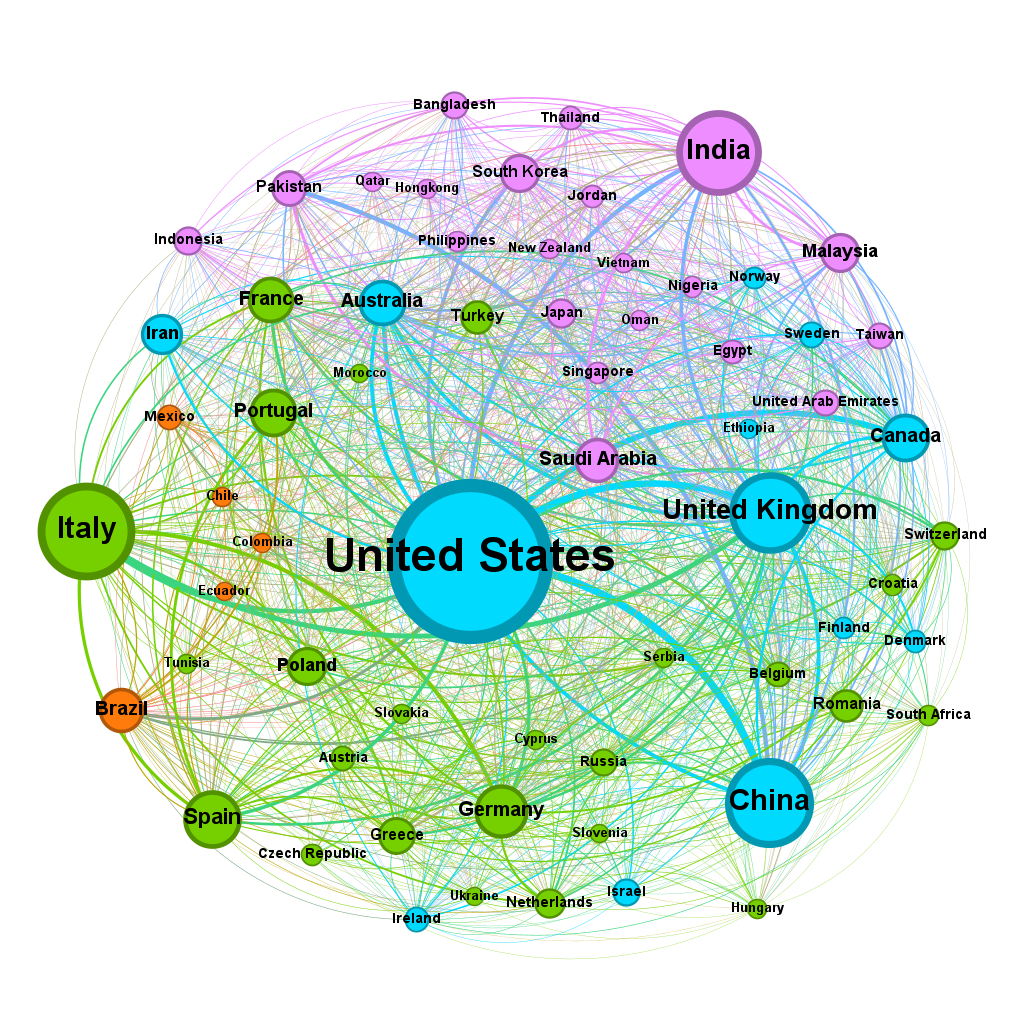}
    \caption{Collaboration network between active countries (produced at least 20 publications).}
    \label{fig:countries-nw22}
\end{figure}

The 20 most active countries contributing to disaster informatics during the investigated period are presented in Figure \ref{fig:top-countries22}, where the contribution of each country is based on the affiliation with the authors of an article. Our results indicate that the U.S., China, and India are among the most active contributors to disaster informatics from January 2020 to September 2022. Particularly, the U.S. leads with the highest number of publications (981). Italy follows with 534, China with 478, India with 453, and the United Kingdom (UK) with 431. The findings in our study align with previous studies that the U.S. and India were the most active countries in disaster informatics between 1960 and 2020 \citep{Ogie2020DisasterIA}, while the U.S. and China were the most active countries in contributing to information behaviors during natural disasters from 2000 to 2021 \citep{Tan_Hao_2022}.

% Ogie and Verstaevel found that the U.S. and India were the most active countries in disaster informatics between 1960 and 2020 \citep{Ogie2020DisasterIA}. Tan and Hao found that the U.S. and China were the most active countries in contributing to information behaviors during natural disasters from 2000 to 2021 \citep{Tan_Hao_2022}.
% These findings support our results, in which the U.S., China, and India are among the most active contributors to disaster informatics from January 2020 to September 2022. Particularly, the U.S. leads with the highest number of publications (981). Italy follows with 534, China with 478, India with 453, and the United Kingdom (UK) with 431. 

The collaborations between countries are visualized in Figure \ref{fig:countries-nw22} using Gephi \citep{ICWSM09154}. The color of a node and its link denotes groups of countries that frequently collaborated during the investigated period. The colors blue, green, pink, and red indicate the presence of 4 groups of countries among the most active ones. To enhance visualization clarity, countries with less than 20 publications were filtered out from this network. 

The U.S. was the largest contributor and the leading country in international collaborations with 110 partner countries. Italy ranked second in both publication output and international partnerships, collaborating with 108 countries. Despite ranking fifth in publication output, the UK was third in international collaborations, partnering with 98 countries. India and China followed closely, collaborating with 96 and 92 countries respectively. This pattern of extensive global collaboration highlights the importance of international partnerships in addressing multifaceted challenges in disaster informatics during the pandemic. 

In terms of groups of countries collaborating, the blue group in the network shows a close-knit circle of collaborators including the U.S., the UK, China, Canada, Australia, Iran, Finland, and Sweden among others. In terms of publication output in this group, the U.S., China, and the UK were the most active ones. In terms of international collaborations, the U.S. (110), UK (98), and China (92) had the highest number of collaborations. %Notably, this group contains many English-speaking countries, which may facilitate their frequent collaborations.

The green group in Figure \ref{fig:countries-nw22} includes predominantly European countries, in which Italy, Spain, Portugal, and Germany are the main contributors. Other members of this group include France, Greece, Russia, Poland, Turkey, Romania, South Africa, the Netherlands, and Austria among others. In terms of international collaborations, Italy (108), Germany (89), and France (89) were at the forefront. %This group primarily consists of European countries, which likely facilitate their frequent collaborations. 

The pink group in Figure \ref{fig:countries-nw22} includes mainly Asian countries, in which India, Saudi Arabia, Malaysia, South Korea, and Pakistan are the main contributors. Other members in this group include the United Arab Emirates, Bangladesh, New Zealand, Japan, Indonesia, the Philippines, Vietnam, and Taiwan among others. In terms of international collaborations, India (96), Japan (83), and Saudi Arabia (80) were at the forefront. 

The orange group includes mostly Latin American countries, in which Brazil (186), Mexico (59), and Chile (32) are the main contributors. Other members, including Peru, Argentina, Uruguay, Venezuela, Costa Rica, and Panama among others were omitted from the network since their number of publications was less than 20. Among these countries, Brazil (85), Mexico (77), and Colombia (75) lead in international collaborations.

Overall, the English-speaking group of countries (the blue group) had the highest international collaborations, followed by the group of European countries (the green group), and the group of Asian countries (the pink group). The Latin American group (the orange group) had the lowest number of collaborations and publications. 

%This group primarily consists of European countries, which likely facilitate their frequent collaborations.

% estimated deaths over 4 million in China (cite non-peer-reviewed article https://www.medrxiv.org/content/10.1101/2023.06.15.23291443v2)
\subsubsection{Institutions}

\begin{figure}
    \centering
    \includegraphics[width=0.75\linewidth]{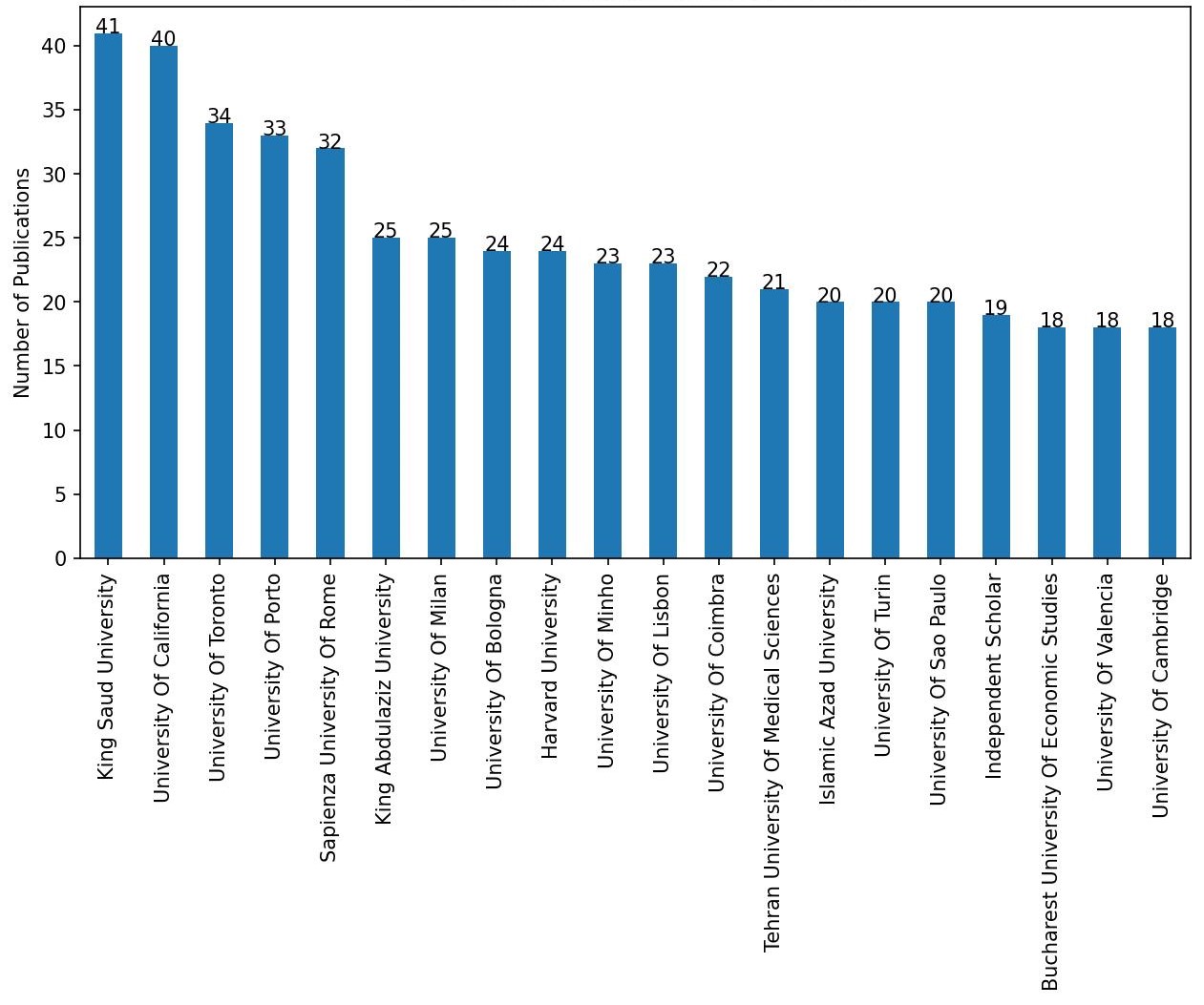}
    \caption{The top 20 institutions.}
    \label{fig:top-institutions22}
\end{figure}

\begin{figure}
    \centering
    \includegraphics[width=0.75\linewidth]{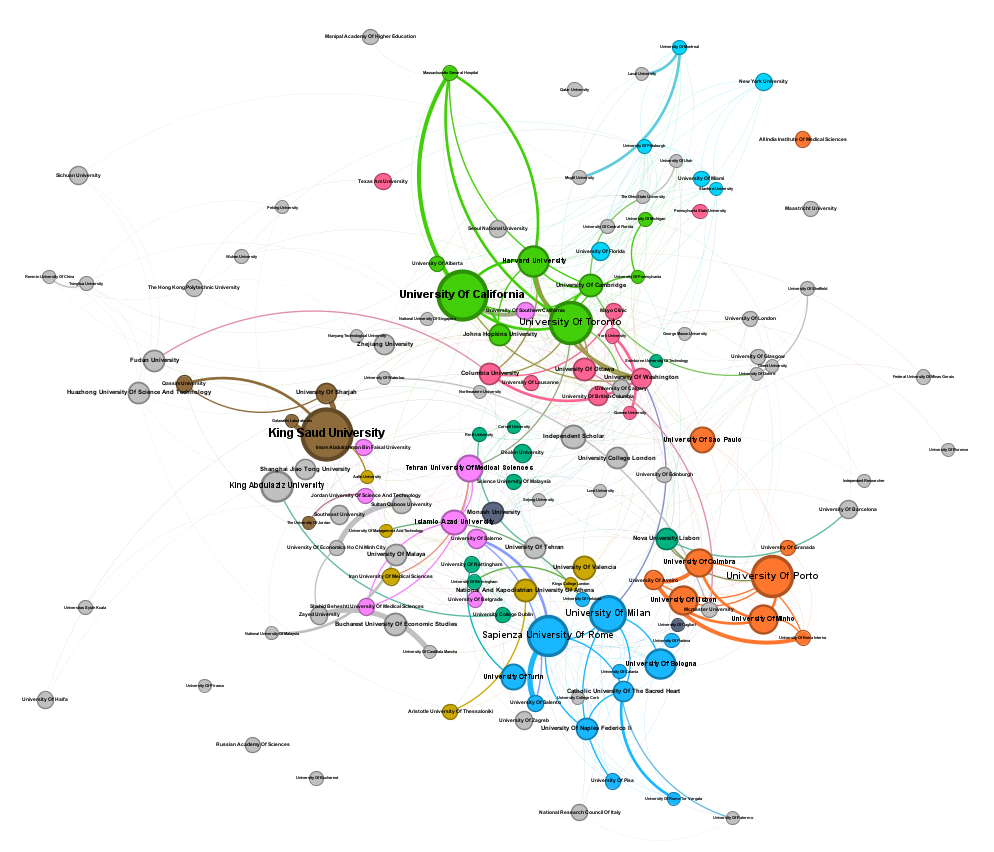}
    \caption{Collaboration network between active institutions (produced at least 10 publications).}
    \label{fig:institutions-nw22}
\end{figure}

The 20 most active institutions are shown in Figure \ref{fig:top-institutions22}.
King Saud University, Saudi Arabia leads with 41 publications, followed by the University of California, U.S. (40 publications), the University of Toronto, Canada (34 publications), the University of Porto, Portugal (33 publications), and Sapienza Univerity of Rome, Italy (32 publications). These institutions are also from the most active countries identified in the previous section. 

The collaborations of these institutions are depicted in the network of Figure \ref{fig:institutions-nw22}. The color of the node and link indicates the groups of institutions that frequently collaborated during the investigated period. Gray nodes represent institutions that either do not belong to any specific group or belong to groups comprising less than 2\% of all institutions. Institutions that produced fewer than 10 publications were excluded from this network. This network features 3 prominent groups, green, blue, and orange. 

The green group includes the University of California (40 publications), the University of Toronto (34 publications), and Harvard University (24 publications) as the main contributors. Institutions from this group are predominantly from English-speaking countries, and most are located in North America. The most frequent collaborations occurred between the University of Toronto and Harvard University, and between the University of California and Massachusetts General Hospital, with each pair producing 4 publications. 

The blue group primarily consists of Italian institutions with the Sapienza University of Rome (32 publications), the University of Milan (29 publications), the University of Bologna (24 publications), and the University of Turin (20 publications) as the main contributors. Notable collaborations within this group include those between the University of Salento and the Sapienza University of Rome, and between the University of Rome Tor Vergata and the Catholic University of Sacred Heart, with each pair producing 5 and 3 publications respectively. 

The orange group features institutions from Portugal, with the University of Porto (33 publications), the University of Minho (23 publications), the University of Lisbon (23 publications), and the University of Coimbra (22 publications) as the main contributors. The most frequent collaborations were between the University of Aveiro and the University of Coimbra, jointly producing 4 publications.

\subsubsection{Authors}
\begin{comment}

\begin{figure}
    \centering
    \includegraphics[width=0.75\linewidth]{Top13-Authors.jpg}
    \caption{The Most Active Authors 2020-Sept.2022}
    \label{fig:top-authors22}
\end{figure}
\end{comment}

%optional
\begin{figure}
    \centering
    \includegraphics[width=0.75\linewidth]{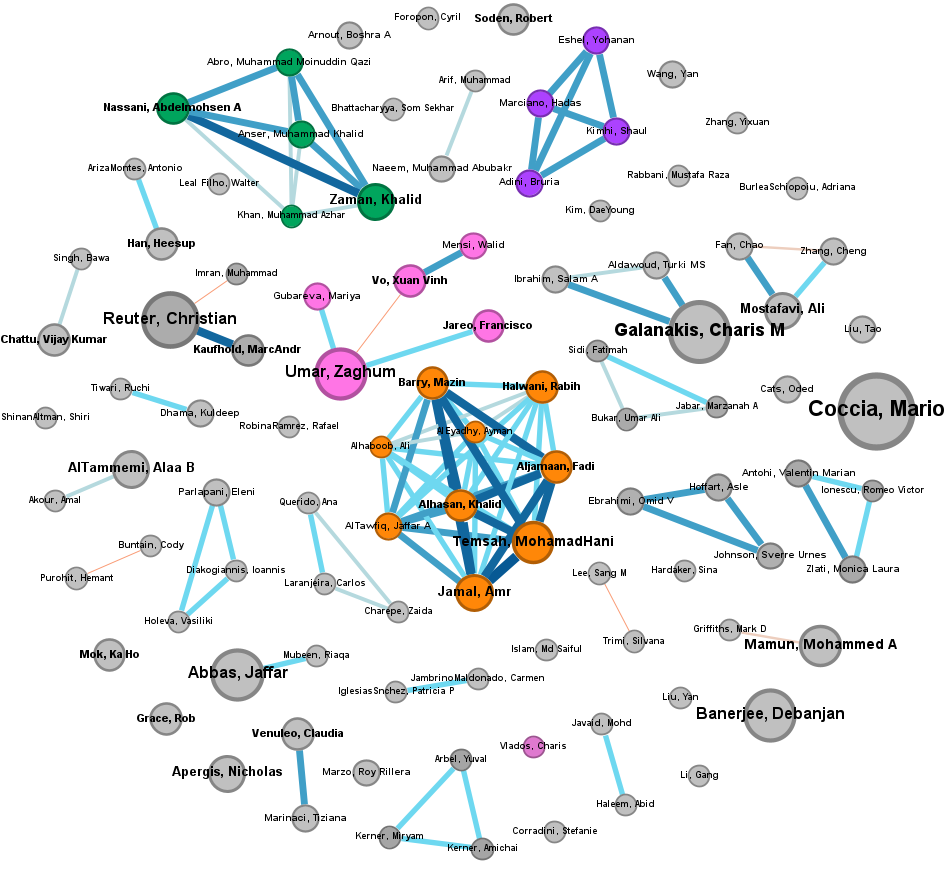}
    \caption{Collaboration network between authors who produced at least four publications.}
    \label{fig:authors-nw22}
\end{figure}

\begin{figure}
    \centering
    \includegraphics[width=0.75\linewidth]{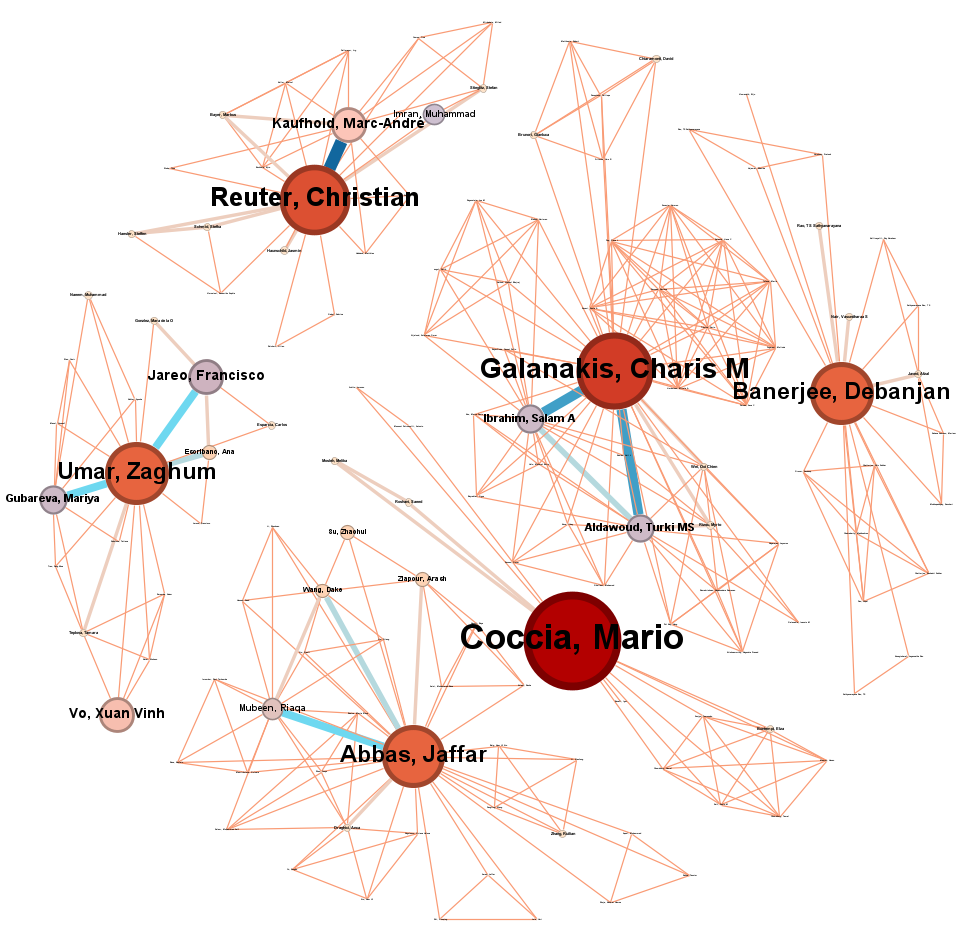}
    \caption{Collaboration network between the six most active authors in disaster informatics.}
    \label{fig:top-authors-nw22}
\end{figure}

\begin{table}
    \centering
    \caption{The most active authors.}
    \label{tab:top-authors22}
    \begin{tabular}{cccc}
        \hline
        \textbf{\# Publications} & \textbf{Author Name} & \textbf{Affiliated Institution} & \textbf{Affiliated Country} \\ \hline
        15 & Mario Coccia  &  National Research Council of Italy & Italy \\ \hline
        12 & Charis M Galanakis  & Galanakis Laboratories & Greece \\ \hline
        11 & Christian Reuter &  Technical University of Darmstadt & Germany \\ \hline
        10 & Zaghum Umar &  Zayed University & United Arab Emirates \\ \hline
        10 & Jaffar Abbas & Shanghai Jiao Tong University & China \\ \hline
        10 & Debanjan Banerjee & National Institute of Mental Health and Neuro Sciences & India \\ \hline
      %  8  & Mohamad-Hani Temsah &  & \\ \hline
      %  8  & Mohammed A Mamun &  & \\ \hline
      %  7  & Amr Jamal &  & \\
      %  7  & Nicholas Apergis &  & \\
      %  7  & Khalid Zaman &  & \\
      %  7  & Ala'a B Al-Tammemi &  & \\
      %  7  & Ali Mostafavi &  & \\
    \end{tabular}
\end{table}

The top 20  authors are shown in Table \ref{tab:top-authors22}. Leading authors are Mario Coccia, Charis Galanakis, Christian Reuter, Zaghum Umar, Jaffar Abbas, and Debanjan Banerjee. Table \ref{tab:top-authors22} also demonstrates that the countries associated with these authors are among the 20 most active, indicating a strong national interest in disaster informatics research. 
% Create a table instead: author name (#publications), institution, and country
%Leading the list is Mario Coccia from the National Research Council of Italy with 15 publications, followed by Charis Galanakis from Galanakis Laboratories, Greece (12 publications), and Christian Reuter from the Technical University of Darmstadt, Germany (11 publications). Other notable contributions include Jaffar Abbas from the Shanghai Jiao Tong University, China, Debanjan Banerjee from the National Institute of Mental Health and Neuro Sciences, India, and Zaghum Umar from Zayed University, United Arab Emirates, each contributing 10 publications. 

%optional
The network illustrated in Figure \ref{fig:authors-nw22} visualizes collaborations between authors, who have produced at least 4 publications each. Each node in the network represents an author, with the node size indicating their publication output. Links between the nodes represent collaboration efforts, which vary in thickness and color, ranging from light orange to dark blue, with thicker, darker blue links indicating more frequent collaborations. The network only includes authors with at least 4 publications, therefore some collaboration links were omitted. The network further highlights 4 prominent groups of authors in orange, pink, green, and purple, which were identified using the function Modularity in Gephi \citep{ICWSM09154}. Notably, the orange group has the most members and collaborations. In this group, the node sizes are similar to each other and so as most of the links between them, suggesting that these authors frequently collaborated in groups. The authors represented in the gray nodes did not collaborate frequently.

Furthermore, the collaborative patterns of the most active authors are visualized in Figure \ref{fig:top-authors-nw22}, which includes authors who have collaborated with the most active authors (red and orange nodes). Each node represents an author. The size and color intensity of the nodes reflect their publication outputs. Larger, darker nodes denote the 6 most active authors. Thick blue links indicate frequent collaborations, while thin orange links represent the contrary.

Among the active authors, a common pattern frequently emerged, with one-time collaborations. All six active authors engaged in this type of collaboration multiple times. However, there is one difference. Four authors formed frequent partnerships with no more than two other authors, while the others did not maintain such frequent partnerships. For instance, Mario Coccia and Debanjan Banerjee did not have any frequent partnerships. In contrast, Christian Reuter frequently partnered with Marc-André Kaufhold, Zaghum Umar with Francisco Jareo and Mariya Gubareva, Charis Galanakis with Salam Ibrahim and Turki Aldawoud, and Jaffar Abbas with Riaqa Mubeen and Dake Wang.

\subsection{Topic Result} %Comparison Results from Different Topic Modeling Approaches}

\subsubsection{Topic Modeling Result}
\label{sec:topic-model-comparison}

\begin{table}
    \centering
    \caption{Performances of four topic modeling methods over different evaluation metrics.}
    \label{tab:topic-model-performance}
    %\begin {tabular}{|c|p{0.15\textwidth}|p{0.15\textwidth}|p{0.15\textwidth}|p{0.15\textwidth}|} \hline
    \begin{tabular}{c>{\centering\arraybackslash}m{0.10\textwidth}>{\centering\arraybackslash}m{0.15\textwidth}>{\centering\arraybackslash}m{0.15\textwidth}>{\centering\arraybackslash}m{0.15\textwidth}} \hline    
         
         Metrics & BERTopic & BERTopic-BERTopic & BERTopic-GloVe & BERTopic-BioMedLM \\ \hline
        Number of Topics & 33 & 50 & 34 & 51\\ \hline 
        Avg. Coherence ($\uparrow$) & \textbf{0.57} & 0.52 & 0.51 & 0.51\\ \hline 
        Perplexity ($\downarrow$) & 1.56 & \textbf{1.27} & 1.28 & 13.01\\ \hline
        Diversity ($\uparrow$) & \textbf{0.63} & 0.47 & 0.44 & 0.47\\ \hline
        Avg. Topic Embedding Similarity ($\downarrow$) & \textbf{0.63} & 0.69 & 0.65 & 0.81\\ \hline 
        Avg. Topic Significance ($\uparrow$) & \textbf{1.28} & 0.97 & 0.79 & 1.02\\ \hline
    \end{tabular}
\end{table}

Prior to conducting topic analysis, we applied four topic models, BERTopic, BERTopic-BERTopic, BERTopic-GloVe, and BERTopic-BioMedLM to the collected literature and evaluated their performances. The best-performing topic model was used to extract the underlying topics.
Unlike conventional topic models, the selected models do not require a specific input of the number of topics; instead, they require an input defining the minimum cluster size needed to form a topic. This approach enables more flexibility and adaptability in identifying topics within the data.

Based on our preliminary experiment results, we selected the minimum cluster size n to be 30 and m to be 15. Table \ref{tab:topic-model-performance} contains the number of topics extracted by each topic model and its performance metrics, including the average coherence, perplexity, diversity, average topic embedding similarity, and average topic significance. 

BERTopic by itself extracted the least number of topics (33 topics), and has the highest average coherence score, diversity score, and average topic significance, along with the lowest topic embedding similarity. This indicates that BERTopic is proficient at generating topics that are logically consistent, diverse, unique, and informative. BERTopic also has a relatively low perplexity score (1.56), indicating the model can form topics with a manageable level of uncertainty, underscoring its effectiveness in topic modeling. 

%We evaluated each topic modeling method via the average coherence score as a proxy. When a topic modeling method performs well, its generated topics are logically consistent and represented by a higher average coherence score.  To get the average coherence score of a topic modeling method, we calculated the coherence score of each topic and average them together to represent the performance of a topic modeling method. 

BERTopic-BERTopic is the second-best-performing model with the lowest perplexity score and the second-best coherence and diversity scores. On the other hand, BERTopic-BioMedLM generated the highest number of topics (51 topics) with the worst coherence, perplexity, and topic-embedding similarity scores. In contrast, BERTopic-GloVe performed worst as it generated the lowest coherence and diversity scores. 

This result indicates that BERTopic is the best-performing model for extracting topics from the collected data. It also indicates that a high number of topics does not necessarily mean improved quality, emphasizing the model's ability to extract meaningful topics without relying on sheer volume.

To proceed with topic analysis, we selected 12 topics generated by BERTopic with the highest topic significance scores for summarization and further analysis. These topics were chosen because they represent the most informative research areas with a strong thematic focus, making them ideally suited for in-depth analysis and interpretation. This selection allows us to pinpoint specific research areas within the field of disaster informatics from January 2020 to September 2022.

\subsubsection{Topic Description Result} %Evaluation of LLM-Generated Topic Descriptions}
\label{sec:topic-description-evaluation}

To generate the descriptions of the selected 12 topics, we used GPT-3.5-turbo and Llama 2 to summarize each topic based on its abstracts, and then they were evaluated by two human evaluators. This evaluation involved examining the consistency between each topic description and the corresponding abstracts, ensuring that the summaries accurately reflected the topics presented in the abstracts. 

The results reflecting unanimous agreement between evaluators, are displayed in Table \ref{tab:topic_description_eval} for the topic descriptions generated by GPT-3.5-turbo and Llama 2 respectively. The table includes the topic ID and the number of documents for each topic, the number of abstracts aligned with the topic description as unanimously rated by both evaluators, the level of comprehensiveness, and Cohen's Kappa - the level of agreement between both evaluators. In the table, the level of comprehensiveness and Cohen's Kappa statistics are compared with each other. Bold values indicate higher values, while values followed by an asterisk (*) indicate a tie. Cohen's Kappa values range from 0 (random agreement) to 1 (absolute agreement), with values of at least moderate agreement (at least 0.40) marked with a \dag\/.

In Table \ref{tab:topic_description_eval}, the level of comprehensiveness, agreed upon by both evaluators, indicates how well a topic aligns with the corresponding abstracts. Specifically, the descriptions generated by GPT-3.5-turbo for 4 topics (Topics 8, 10, 11, and 12) showed better consistency with the corresponding abstracts, compared to those generated by Llama 2. However, the descriptions generated by Llama 2 for 5 topics (Topics 1, 2, 6, 7, and 9) have higher levels of comprehensiveness, compared to those generated by GPT-3.5-turbo.
% This value ranges from 0.43 to 0.81. The descriptions generated by GPT-3.5-turbo and Llama 2 for 3 topics (Topics 3, 4, and 5) achieved the same level of comprehensiveness. The inter-rater agreement level between both evaluators for GPT-3.5-generated topic descriptions ranges from 0.23 (slight agreement) to 1.0 (absolute agreement).

In terms of the inter-agreement level between two evaluators, there is at least substantial agreement (a Cohen's Kappa value of at least 0.40) across 9 descriptions generated by GPT-3.5-turbo. However, there is substantial to perfect agreement among both evaluators (a Cohen's Kappa value ranging from 0.40 to 1) in the evaluation of Llama 2-generated descriptions for 11 out of 12 topics (except Topic 8).

%In Table \ref{tab:llama2-topic-description-eval}, the level of comprehensiveness, agreed upon by both evaluators, ranges from 0.48 to 0.81. Specifically, . 
%The agreement level between both evaluators for LLama 2-generated descriptions ranges from 0.20 (slight agreement) to 1.0 (perfect agreement).  

% Comparison between GPT-3.5 and Llama2
The topic descriptions generated by Llama 2 show higher levels of comprehensiveness compared to those generated by GPT-3.5-turbo, indicating that these descriptions are better aligned with the topic's abstracts. Additionally, the high level of agreement between both evaluators across nearly all topics (except Topic 8) when assessing Llama 2-generated descriptions underscores the reliability and consistency of their assessments. 

Based on the evaluation results, the descriptions generated by Llama 2 for the selected 12 topics will be used for the topic discussion below.

\begin{table}[pos=hbt]
%\begin{table}[h]
%{>{\centering\arraybackslash}m{0.5cm}>{\centering\arraybackslash}m{0.5cm}Y>{\centering\arraybackslash}m{2cm}>{\centering\arraybackslash}m{1cm}}
    \centering
    \caption{The mutually agreed evaluations of GPT-3.5-turbo- and Llama 2-generated topic descriptions. (\textbf{Bold} indicates higher, * indicates a tie, and \dag \space indicates substantial agreement in Cohen's Kappa).}
    \label{tab:topic_description_eval}
    \begin{tabular}{cccccccc}
    \toprule
        \multirow{2}{*}{\centering Topic ID} & \multirow{2}{*}{\centering \# Docs} & \multicolumn{3}{c}{GPT-3.5-turbo} & \multicolumn{3}{c}{Llama 2}\\
        \cmidrule(lr){3-5} \cmidrule(lr){6-8}
         & & \makecell{\# Aligned\\Abstracts} & \makecell{Level of\\Comprehensiveness} & \makecell{Cohen's \\Kappa} & 
        \makecell{\#Aligned\\Abstracts} & \makecell{Level of\\Comprehensiveness} & \makecell{Cohen's \\Kappa}\\ \midrule
         1 & 36 & 24 & \textbf{0.67}& \textbf{0.4} & 22 & 0.61 & 0.20 \\
         2 & 32 & 20 & \textbf{0.62}& $\textbf{0.86}^\dag$ & 16& 0.5 &  $0.62^\dag$ \\
         3 & 37 & 30 & $0.81^*$  & $1.00^*$$^\dag$+& 30  & $0.81^*$ &    $1.00^*$$^\dag$    \\
         4 & 37 & 20 & $0.54^*$  & $0.78^\dag$     & 20  & $0.54^*$ &    $0.72^\dag$ \\ 
         5 & 31 & 14 & 0.45      & $0.81^*$$^\dag$ & 15  & \textbf{0.48} &    $0.81^*$$^\dag$ \\
         6 & 38 & 21 & \textbf{0.55}& $\textbf{0.66}^\dag$& 18 & 0.47 &  $0.53^\dag$ \\
         7 & 42 & 18 & 0.43      & $\textbf{0.63}^\dag$  & 20 & \textbf{0.48}     &    $0.56^\dag$ \\
         8 & 49 & 28 & 0.57      & $0.54^\dag$ & 30 & \textbf{0.61} &    $\textbf{0.56}^\dag$\\
         9 & 44 & 19 & 0.43      & 0.23 &  21 & \textbf{0.48} &    $\textbf{0.42}^\dag$ \\
         10 & 61 & 30 & $0.49^*$ & 0.4 & 30 & $0.49^*$ &    $\textbf{0.45}^\dag$  \\ 
         11 & 53 & 26 & 0.49     & $0.59^\dag$ & 31 & \textbf{0.58} &    $\textbf{0.71}^\dag$ \\
         12 & 61 & 34 & \textbf{0.56}& $\textbf{0.44}^\dag$ & 30 & 0.49 &  $\textbf{0.52}^\dag$ \\ 
         %Biomedical Science & Cell (1974 - 2020) & 15,204 & Cell 2021 &  447 \\ 
        \bottomrule
    \end{tabular}
    \label{dataset}    
\end{table}
%\end{landscape}

\begin{comment}
\begin{table}[h]
\centering
\caption{The mutually agreed evaluations of Llama 2-generated topic descriptions by both evaluators and the levels of agreement. (\textbf{Bold} indicates higher, * indicates a tie, and \dag indicates substantial agreement in Cohen's Kappa).}
\label{tab:llama2-topic-description-eval}
\begin{tabular}{|c|c|c|c|c|} \hline       %{|l|l|l|l|l|} 
\makecell{Topic ID} & 
\makecell{\# Docs} & 
\makecell{\# Abstracts \\ Aligned with a \\ Topic Description} & 
\makecell{Probability of \\ Abstracts Aligned \\ with a Topic Description} & 
\makecell{Cohen's Kappa \\ Statistic}  \\ \hline
1  & 36 & 22 & 0.61 & 0.20 \\ \hline 
2  & 32 & 16 & 0.5 &  $0.62^\dag$ \\ \hline 
3 & 37 & 30 & $0.81^*$ &    $1.00^*$$^\dag$    \\ \hline 
4 & 37 & 20 & $0.54^*$ &    $0.72^\dag$ \\ \hline 
5  & 31 & 15 & \textbf{0.48}     &    $0.81^*$$^\dag$ \\ \hline 
6  & 38 & 18 & 0.47 &  $0.53^\dag$ \\ \hline 
7  & 42 & 20 & \textbf{0.48}     &    $0.56^\dag$ \\ \hline 
8 & 49 & 30 & \textbf{0.61}     &    $\textbf{0.56}^\dag$\\ \hline
9 & 44 & 21 & \textbf{0.48}     &    $\textbf{0.42}^\dag$       \\ \hline 
10 & 61 & 30 & $0.49^*$ &    $\textbf{0.45}^\dag$         \\ \hline 
11 & 53 & 31 & \textbf{0.58}     &    $\textbf{0.71}^\dag$ \\ \hline 
12 & 61 & 30 & 0.49 &  $\textbf{0.52}^\dag$ \\ \hline
\end{tabular}
\end{table}
\end{comment}

\subsection{The 12 Most Significant Topics} %Discussion of the Top 12 Topics} 
\label{sec:topic_discussion}
This section briefly describes the 12 topics with the highest significance scores. Table \ref{tab:topicID-title} lists these topics sorted decreasingly by their significance scores, along with the number of documents each topic contains, the titles generated by Llama 2, the topic keywords generated by KeyBERT \citep{grootendorst2020keybert}, and their significance scores. For more detailed descriptions of the most significant topics, see Appendix \ref{sec:appendix_topic_description}.

\newcolumntype{Y}{>{\centering\arraybackslash}m{4cm}}
\newcolumntype{Z}{>{\centering\arraybackslash}X}
\renewcommand{\tabularxcolumn}[1]{>{\arraybackslash}m{#1}}

\begin{longtable}{>{\centering\arraybackslash}m{0.8cm}>{\centering\arraybackslash}m{1cm}Y>{\centering\arraybackslash}m{5.6cm}>{\centering\arraybackslash}m{1.6cm}}
\caption{The 12 most significant topics and topic titles.} 
\label{tab:topicID-title} \\
\hline  
\textbf{Topic ID} & \makecell{\textbf{\# Docs}} & \textbf{Topic Title} & \textbf{Topic Keywords} & \makecell{\textbf{Significance} \\ \textbf{Score}} \\ \hline    
\endfirsthead
\caption[]{(continued)} \\
\hline  
\textbf{Topic ID} & \makecell{\textbf{\# Docs}} & \textbf{Topic Title} & \textbf{Topic Keywords} & \makecell{\textbf{Significance} \\ \textbf{Score}} \\  \hline    
\endhead
\hline \endfoot

1 & 36 & Impact of COVID-19 on International Trade & trade, export, international, policy, international trade, country, trade policy, product, crisis, COVID-19 & 1.98 \\ \hline

2 & 32 & Burnout Among Healthcare Professionals During the COVID-19 Pandemic & burnout, work, relate, work relate, job, nurse, professional, staff, physical, factor & 1.9 \\ \hline

3 & 37 & Medical Technology and Additive Manufacturing during COVID-19 Pandemic & print, equipment, medical, device, additive manufacture, printing, technology, COVID-19, manufacturing, medical device & 1.85 \\ \hline

4 & 37 & Waste Management and Environmental Impacts of COVID-19 & waste, plastic waste, waste management, management, environmental, plastic, face mask, COVID-19, generation, environment & 1.76 \\ \hline

5 & 31 & Community Pharmacy Practice During COVID-19 & pharmacy, pharmacist, community pharmacist, community pharmacy, COVID-19, patient, medication, service, pandemic, pharmaceutical & 1.75 \\ \hline

6 & 38 & Genomics and Variants of SARS-CoV & SARS-CoV, variant, sequence, genome, mutation, lineage, protein, genetic, epitope, viral & 1.74 \\ \hline

7 & 42 & The Role of Nutrition in the Immune System during COVID-19 & vitamin, immune, COVID-19, antioxidant, extract, disease, food, infection, property, potential & 1.67 \\ \hline

8 & 49 & COVID-19 Pandemic's Impact on Dental Education and Practice & 
dental, dentist, COVID-19, student, patient, care, dental education, practice, pandemic, dental professional & 1.64 \\ \hline

9 & 44 & Impact of COVID-19 Pandemic on Pregnant Women and Maternal Health & pregnant woman, pregnancy, COVID-19, woman, abortion, pandemic, care, maternal, breastfeed, health & 1.58 \\ \hline

10 & 61 & Consumer Behavior and Trust During the COVID-19 Pandemic & consumer, trust, customer, pandemic, brand, COVID-19, service, perceive, crisis, intention & 1.58 \\ \hline

11 & 53 & Virus Inactivation and Detection & virus, SARS-CoV, inactivation, surface, detection, UV, material, food, base, antimicrobial & 1.55 \\ \hline

12 & 61 & The Role of Technology in the COVID-19 Pandemic & technology, app, contact trace, user, privacy, digital, COVID-19, data, social, pandemic & 1.5 \\ \hline

\end{longtable}

\subsection{Topic-Wise Bibliometric Result}
%Authorship Productivity and Publication Patterns in the Top 12 Topics %based on the 12 topics
\label{sec:bibliometric_analysis_12topics}
This section includes the result of our bibliometric analysis of the leading countries, institutions, and authors and their collaborations across the 12 most significant topics.
Table \ref{tab:authorship_12topics} in the Appendix \ref{sec:appendix_authorships} shows the names of the most active authors, institutions, and countries in each of the top 12 topics along with the number of publications. In this study, active authors are those who produced the highest number of publications. Active institutions and active countries are defined similarly. 

%In Table \ref{tab:authorship_12topics}, * indicates that in a given topic, many authors (or institutions) produced the same number of publications and there are no authors (or institutions) with the highest number of publications. For instance, in Topic 3, all authors produced only 1 publication, hence there is an * followed by number 1. Similarly, in the same topic, all institutions produced only 1 publications, hence the * and number 1. Additionally, \dag indicates that in Topic 11, there are 20 authors with only publications each.

\subsubsection{The Leading Countries}
% Which countries/groups of countries work on which topics?

\begin{figure}
    \centering
    \includegraphics[width=0.75\linewidth]{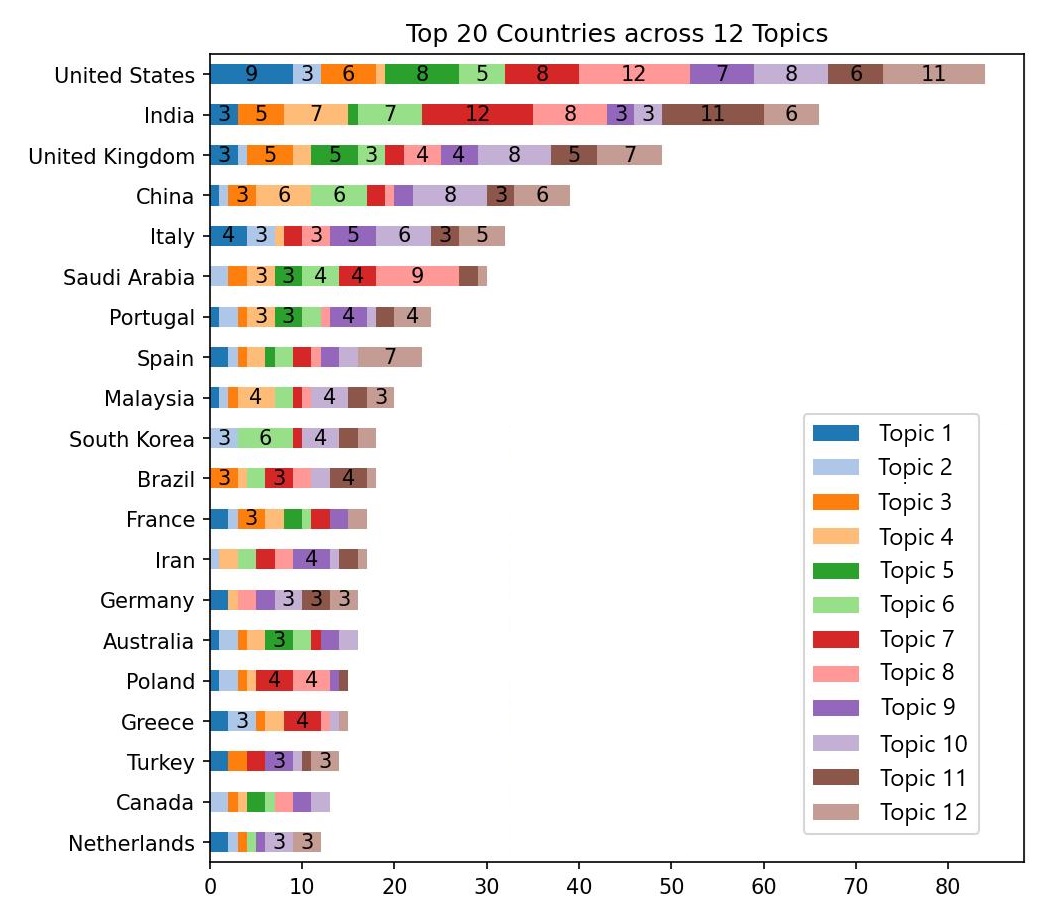}
    \caption{The leading countries in contributing to the 12 topics.}
    \label{fig:top-countries-12Topics}
\end{figure}

\begin{figure}
    \centering
    \includegraphics[width=0.75\linewidth]{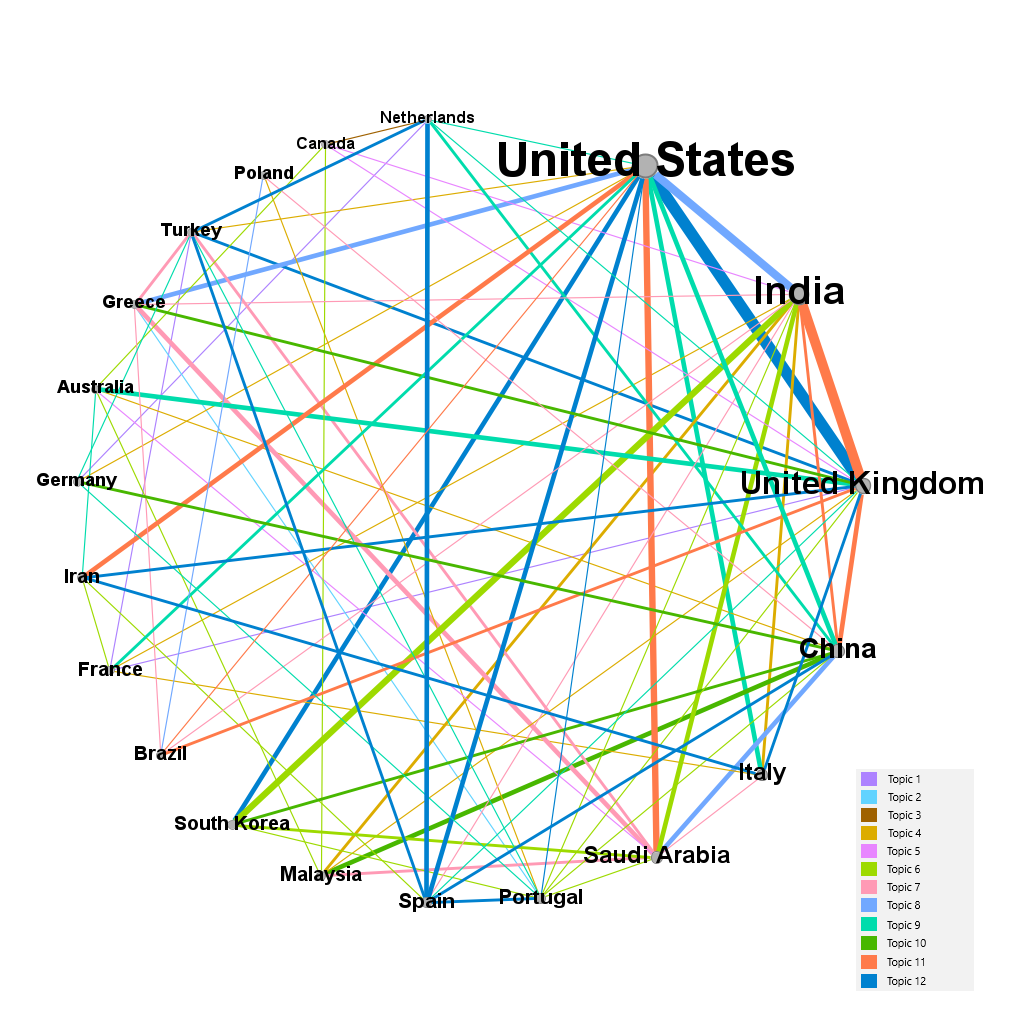} %Top9-countries-nw-12topics.png}
    \caption{Collaboration network between the most active countries across 12 topics (The color of the links corresponds to the specific topic on which the countries collaborated).}
    \label{fig:top-countries-nw-12Topics}
\end{figure}

Figure \ref{fig:top-countries-12Topics} shows the 20 most active countries contributing to the top 12 topics. The number of publications produced by each country is color-coded by topic, with the exact number of publications annotated, provided that it is at least 3. The collaboration network between the top 20 countries across the 12 topics is visualized in Figure \ref{fig:top-countries-nw-12Topics}. 

The U.S. is the most active country with 84 publications across 12 topics, with a greater interest in Topics 8, 12, 1, 5, 7, 10, and 9 respectively (see Table \ref{tab:topicID-title} for topic names), as indicated in Figure \ref{fig:top-countries-12Topics}.
%include a paragraph covering the topic names.
Along with Italy, South Korea, and Greece, the U.S. was among the few at the forefront of investigating burnout and mental health challenges in healthcare professionals during the pandemic (Topic 2). It seems the significant strain on the U.S. healthcare system during the COVID-19 pandemic intensified the country's research interest in understanding and addressing mental health impacts on healthcare workers, emphasizing the importance of maintaining their well-being during global health crises. 

Across the 12 topics, the U.S. partnered with the UK (7 times), India (5 times), and Spain, Saudi Arabia, Italy, and China (3 times each).
Specifically, the U.S. collaborated with the UK, Spain, South Korea, and Portugal on Topic 12, with Greece and India on Topic 8, with Iran, Saudi Arabia, and Brazil on Topic 11, and with China, Italy, France, and Netherlands on Topic 9, based on Figure \ref{fig:top-countries-nw-12Topics}.

India is the second most active country with 66 publications across 11 topics with more interest in Topics 7, 11, 8, 4, and 6 respectively (see Table \ref{tab:topicID-title} for topic names), as indicated in Figure \ref{fig:top-countries-12Topics}.
These contributions highlight India's focus on investigating the characteristics of SARS-CoV and its variants while developing approaches through nutrition, medicine, and technology to mitigate them. Additionally, India's long-standing waste management challenges, compounded by the environmental impacts of the pandemic, enhanced this country's research effort in developing effective waste management solutions during this period. Additionally, India was among the few countries with a strong research interest in the role of nutrition and immune function, as well as the impact of the pandemic on dental practice and education.

India collaborated with the U.S. (5 times), the UK (6 times), Saudi Arabia (3 times), and China, Italy, and Malaysia (2 times each) across 12 topics. India had a strong partnership with the U.S. in investigating Topic 8, a frequent partnership with the UK, and an occasional partnership with China in Topic 11, and collaborated with South Korea, Saudi Arabia, and Portugal in Topic 6.

The UK is the third leading country with 49 publications across 12 topics, with a greater interest in Topics 10 and 12 respectively based on Figure \ref{fig:top-countries-12Topics}. With multiple national lockdowns and local lockdowns enforced between 2020 and 2021, the UK experienced a change in the consumption of alcohol, tobacco, food, and others \citep{Naughton_Ward_et_al_2021, Inman_2020}.
This explains the country's research interest in investigating the shift in consumer behaviors during the pandemic, such as grocery shopping patterns, health behaviors, and eating patterns \citep{Ogundijo_Tas_Onarinde_2021, Thompson_Hamilton_et_al_2022}. The UK also expressed a research interest in exploring the use of technology and 3D printing to detect, inactivate, and mitigate the spread of COVID-19, as well as investigating the challenges that community pharmacists faced during the pandemic.

Some of the UK's notable collaborations were with the U.S. (7 times), India (6 times), and China (3 times). The UK had a strong partnership with India and occasional partnerships with China and Brazil in investigating Topic 11. Also, the UK frequently worked with Australia on Topic 9 and partnered with the U.S. and occasionally with Turkey, Iran, and Italy on Topic 12.

We also noticed that many of the most active countries (13 countries) were engaged in Topic 6, especially South Korea, Saudi Arabia, India, and Portugal. Additionally, half of the most active countries were interested and collaborated on Topic 12.

\subsubsection*{The Leading Institutions}

\begin{figure}
    \centering
    \includegraphics[width=0.75\linewidth]{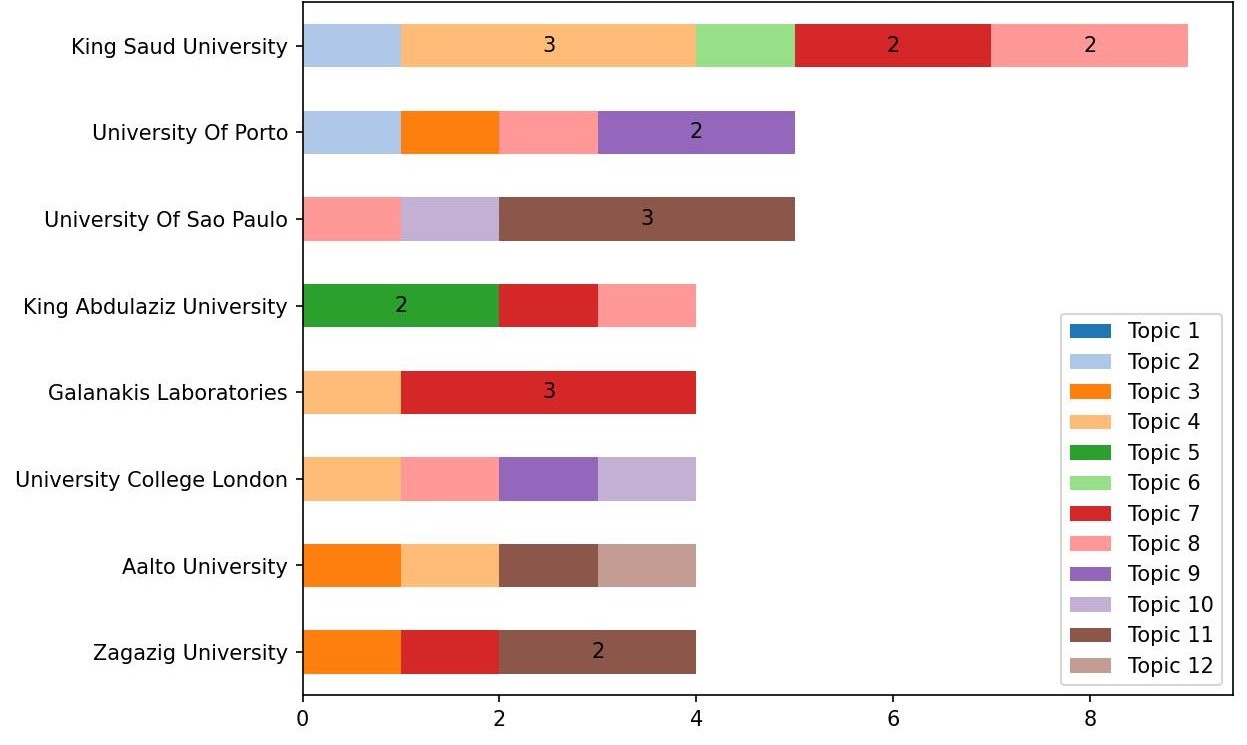}
    \caption{The most active institutions in contributing to the 12 topics.}
    \label{fig:top-institutions-12Topic}
\end{figure}

The 12 most active institutions are visualized in Figure \ref{fig:top-institutions-12Topic}. The number of publications of at least 2 is annotated in the part of a bar corresponding to a particular topic. Based on the publication output in this figure, Topics 4, 5, 7, 8, 9, and 11 were popular among the most active institutions, with each topic garnering at least 2 publications. On the other hand, Topics 1, 2, 6, 10, and 12 were less popular during the investigated period, which may be because research in these areas required more time, or because they were relatively novel at the time.

King Saud University is the most active institution (9 publications) across five topics, focusing on Topics 4, 7, and 8.
The University of Porto and the University of S\~ao Paulo were the second most active institutions, each with 5 publications. While the University of Porto was interested in Topics 2, 3, 8, and 9, the Univerity of S\~ao Paulo was interested in Topics 4, 10, and especially 11. 

\begin{figure}
    \centering
    \includegraphics[width=0.75\linewidth]{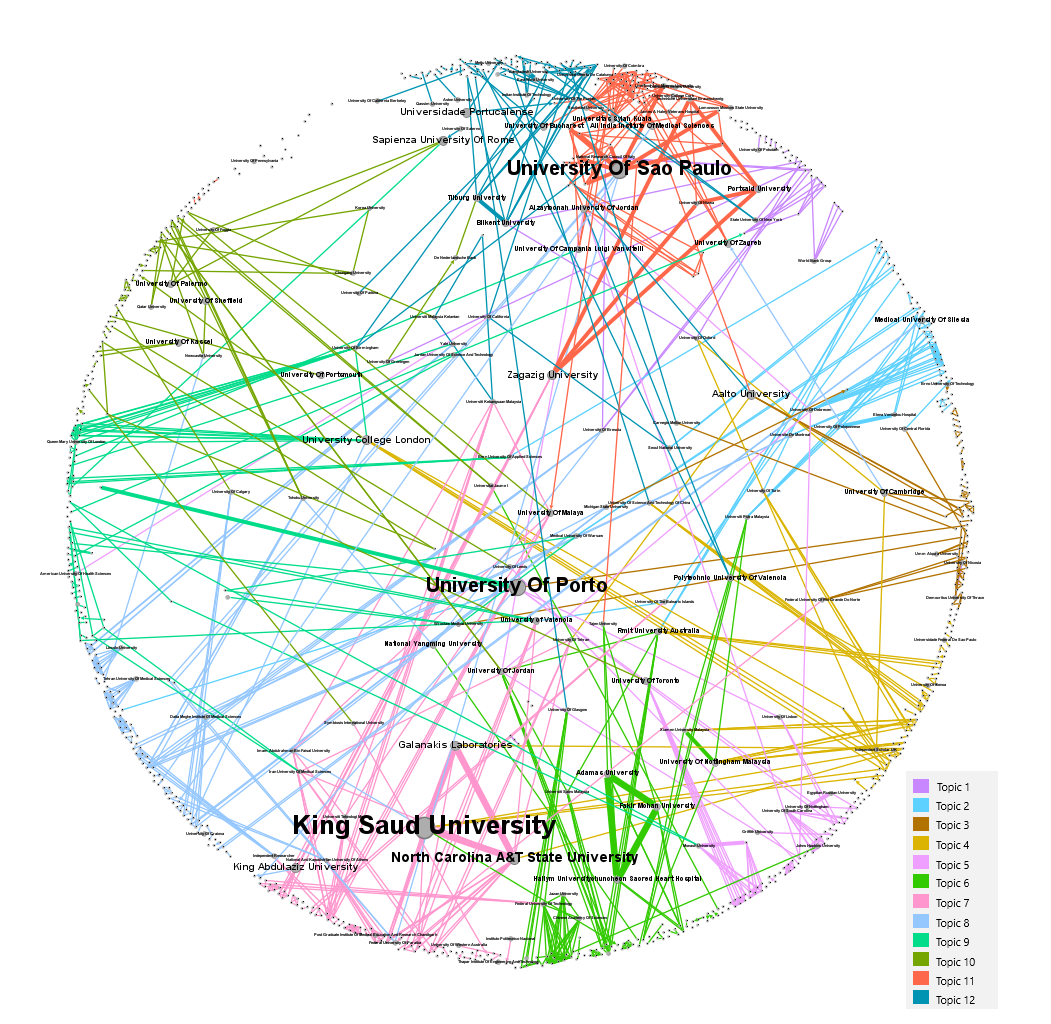}
    \caption{Collaboration network between institutions and their countries (The color of the links denotes the topic that both institutions collaborated on).}
    \label{fig:institutions-nw-12Topics}
\end{figure}

Figure \ref{fig:institutions-nw-12Topics} visualizes the collaboration network between all institutions across 12 topics. Many collaborations occurred on Topics 4, 6, 7, 8, 9, and 11. Particularly, the group of King Saud University (Saudi Arabia), Galanakis Laboratories (Greece), and North Carolina Agricultural and Technical University (U.S.) frequently collaborated on Topic 7. The group of Hallym University Chuncheon Sacred Heart Hospital (South Korea), Fakir Mohan University (India), and Adamas University (India) frequently collaborated on Topic 6. The group of Port Said University (Egypt) and Zagazig University (Egypt) frequently collaborated on Topic 11. In contrast, there were fewer collaborations between institutions on Topics 1, 3, and 5. Given the extensive links in Figure \ref{fig:institutions-nw-12Topics}, there are institutions that investigated multiple topics. For instance, the University of Porto worked together with other institutions on Topics 9, 5, 3, and the University College London worked with others on Topics 9, 10, 8 and 4.

\subsubsection{The Leading Authors}
% Which authors/groups of authors work on which topics?
\begin{figure}
    \centering
    \includegraphics[width=0.75\linewidth]{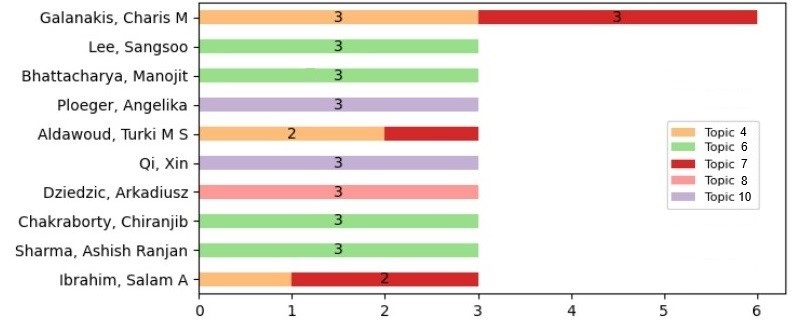}
    \caption{The most active authors in contributing to the top 12 topics.}
    \label{fig:top-authors-12Topic}
\end{figure}

Figure \ref{fig:top-authors-12Topic} indicates the top 10 authors contributing to the top 12 topics. The bar color indicates the topic that the authors contributed, and the number indicates the number of publications on a corresponding topic. 
The legend in Figure \ref{fig:top-authors-12Topic} indicates that most active authors focused on investigating Topics 4, 6, 7, 8, and 10.
Charis Galanakis was the leading author across the top 12 Topics, with 3 publications on Topics 4 and 7. Turki Aldawoud and Salam Ibrahim were also among the most active authors contributing to Topics 4 and 7. Both were frequent collaborators with Charis Galanakis, which explains their interest in the same topics.

\begin{figure}
    \centering
    \includegraphics[width=0.75\linewidth]{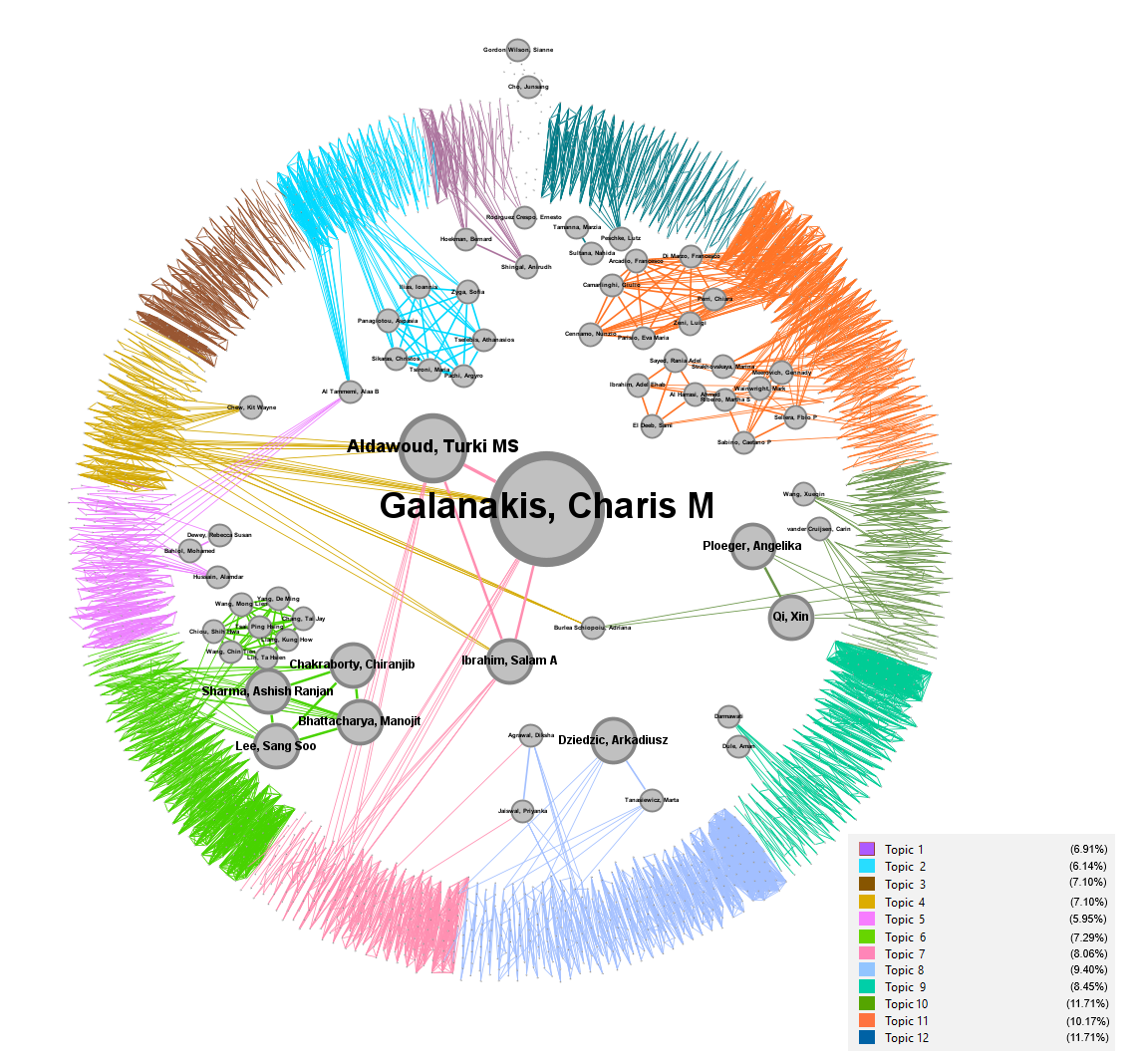}
    \caption{Collaboration network between authors in the top 12 topics.}
    \label{fig:authors-nw-12Topics}
\end{figure}

The collaborations between authors across 12 topics are visualized in Figure \ref{fig:authors-nw-12Topics}. The node's size represents the number of publications of an author. The color of a link denotes a topic on which both authors collaborated. The figure legend shows the color of each topic and the percentage of documents across 12 topics. 

Most authors produced only one publication on one topic. Particularly, Angelika Ploeger is one of the most active authors in Topic 10, Arkadiusz Dziedzic in Topic 8, and Sang Soo Lee in Topic 6. Additionally, authors who worked on multiple topics include Charis Galanakis and Turki Aldawoud (Topics 4 and 7), Adriana Schiopoiu Burlea (Topics 4 and 10), Ala'a B. Al-Tammemi (Topics 2 and 5), and Diksha Agrawal and Priyanka Jaiswal (Topics 7 and 8).

In terms of collaborations, Topics 6, 8, 9, and 11 exhibit denser connections, indicating more extensive collaborations within each of these topics. Additionally, Figure \ref{fig:authors-nw-12Topics} reveals numerous groups of active authors in each topic.
Notable among these are two groups of authors on Topic 6, a group of 4 authors on Topic 8, and two groups of authors on Topic 11.

\section{Discussion}
\label{sec:discussion}

In this section, we present a discussion of our results, including the bibliometric analysis from the collected data, the topic models, the most significant topics, and the bibliometric analysis from the most significant topics.
%in the patterns of publications and collaborations between countries, institutions, and authors from the collected data.
\subsection{Overall Discussion}

\subsubsection{ The overall patterns of publications and collaborations} %RQ1: What are the patterns of publication and collaborations between countries, institutions, and authors in disaster informatics from January 2020 to September 2022?}

The bibliometric analysis across the processed documents showed that the U.S., Italy, China, India, and the UK were the most active countries in contributing and collaborating in the disaster informatics field. These countries were among the most affected by the pandemic by the COVID-19 pandemic from 2020 to 2022, according to Our World in Data\footnote{\url{https://ourworldindata.org/covid-deaths}}.
%The COVID-19 death toll in the U.S. was nearly 1.08 million, in Italy was over 184,000, in China was almost 39,000, in India was over 530,000, and in the UK was over 215,345. 
This pattern suggests that the urgency and severe challenges posed by the COVID-19 pandemic likely motivated these countries to reinforce their scientific and medical research dealing with disaster informatics.
%As these countries were among the most affected by the pandemic, it indicates that challenges posed by the COVID-19 pandemic likely motivated these countries to reinforce their research effort in this field.

Four groups of countries that frequently collaborated were identified as the English-speaking group, the European group, the Asian group, and the Latin American group. It seems that geographical proximity and/or sharing a common language can influence collaborations between countries. The extensive international collaborations reflect a global recognition of the nature of disaster and crisis-related challenges. This underscores the importance of diverse perspectives and knowledge exchange in developing effective solutions to global crises.

In terms of institutions, King Saud University (Saudi Arabia), the University of California (U.S.), the University of Toronto (Canada), the University of Porto (Portugal), and Sapienza Univerity of Rome (Italy) were the most active institutions in publication output. Three groups of institutions that frequently collaborated were identified as the English-speaking group, the Italian group, and the Portugal group. 
These patterns suggest that institutional collaborations are influenced by geographical proximity and common language. This is evident in the formation of research groups among institutions within the same country. Similarly, collaborations among North American universities highlight that a shared language and regional proximity can foster academic partnerships.  
% (the University of California, the University of Toronto, and Harvard University).   (the Sapienza University of Rome, the University of Milan, the University of Bologna, and the University of Turin). (the University of Porto, the University of Minho, the University of Lisbon, and the University of Coimbra) This pattern of collaboration highlights that countries and institutions within a region or share the same language tend to collaborate. , such as the Italian universities (the blue group) and the Portuguese universities (the orange group) (the green group)

Mario Coccia (Italy), Charis Galanakis (Greece), and Christian Reuter (Germany) were the most active authors. In terms of author partnerships, while most authors have extensive networks of occasional collaborations, many active authors tend to form close partnerships with just one or two key partners. This type of collaboration may reflect strategic choices in maintaining focused and productive research relationships within their areas.

\subsubsection{Effective topic modeling and description methods}%RQ2: What are the most significant topics investigated in disaster informatics from January 2020 to September 2022?}
%RQ2
In terms of topic modeling, BERTopic generated the most optimal topics based on coherence, diversity, topic embedding similarity, and average topic significance, and this suggests that this model is sufficient for identifying topics. Existing topic models like BERTopic can be further enhanced with appropriate embedding, reducing the need to develop new models. 

Regarding topic description, given the higher level of comprehensiveness, Llama 2 was better at describing the top 12 Topics than GPT-3.5-turbo. This suggests that LLMs can be incorporated into the topic modeling process to interpret the extracted topics, offering more contextual insights than relying solely on topic keywords.

\subsubsection{Most significant topics in disaster informatics}
%For further analysis, we selected the 12 topics, which were generated by BERTopic and had the highest Topic Significance score. 
The top 12 topics address a wide range of issues that include a crucial aspect of the societal and healthcare challenges posed by the COVID-19 pandemic. These include the impact of the pandemic on international trade; burnout among healthcare professionals; technological solutions to medical supply shortages; waste management and the environmental impacts during the pandemic; community pharmacy practice during COVID-19; research on genomics and variants of SARS-CoV-2; the influence of nutrition on the immune system; adaptations in dental education and practice; the impact of COVID-19 on maternal health; the shifts in consumer trust and behaviors; advancements in virus detection and inactivation techniques; and, the applications of technology in combating the COVID-19 pandemic. 

These topics are evidence that the field of disaster informatics has evolved from a narrow focus on natural disasters with limited attention to public health, to a more comprehensive approach that addresses a range of pandemic-related challenges, integrating with public health, technological solutions, environmental impacts, and beyond.

\subsubsection{The patterns of publications and collaborations across the most significant topics} %RQ3: What are the patterns of publication and collaborations between countries, institutions, and authors across the most significant topics from January 2020 to September 2022?}
% RQ3
%bibliometric analysis results across the most significant topics showed that the 
The U.S., India, and the UK were the most active contributors, with each having specific interests in various topics. The U.S. expressed a special interest in the topics related to the pandemic's impacts on dental education and practice, international trade, community pharmacy practice, consumer behavior and trust, maternal health, use of technology, and the role of nutrition in the immune system during COVID-19. 
India prioritized topics related to the role of nutrition in the immune system, virus inactivation and detection, genomics and variants of SARS-CoV, and the pandemic's impact on dental education and practice, and waste management and the environment. 
The UK focused on topics involved consumer behavior and trust, and the use of technology. In terms of collaborations, the U.S., UK, and India had a close partnership with each other, with each pair of countries working together on different topics. 
Additionally, these three countries have a common interest in topics related to the use of technology, virus inactivation and detection, and medical technology and additive manufacturing during COVID-19.

%12, 11, 3
%1, 3, 6, 8, 9, 10, 11, and 12.
%Topics 8, 12, 1, 5, 7, 10, and 9 respectively, which involve. India 7, 11, 8, 4, and 6. 10 and 12 respectively

In terms of the most active institutions, King Saud University, the University of Porto, and the University of Sao Paulo were the most active, contributing to 5, 4, and 3 of the most significant topics respectively. More institutional collaborations occurred on the topics of waste management and the environmental impacts of COVID-19, genomics and variants of SARS-CoV, and the role of nutrition in the immune system, while fewer collaborations on the topics of the impact of COVID-19 on international trade, medical technology and additive manufacturing, and community pharmacy practice during COVID-19. 

 %For instance, the group of King Saud University, Galanakis Laboratories, and North Carolina Agricultural and Technical University frequently collaborated on Topic 7. 
 
%(4, 6, 7)  (1, 3, and 5). In terms of collaborations, many collaborations occured in the topics of (7, 6, and 11).
Charis Galanakis is the most prolific author across the most significant topics, and his research interest is in waste management and the environmental impacts of COVID-19, and the role of nutrition in the immune system. There are more collaborations between authors in the topics of genomics and variants of SARS-CoV, virus inactivation and detection, and the impacts of COVID-19 on dental education and practice, and maternal health, which may be attributed to the complex nature of these topics. The pattern of collaboration between authors suggests that authors tend to specialize in a specific topic or two. This is different from the collaboration pattern between institutions, in which an institution has more diverse interests in various topics. 
%6, 8, 9, and 11. Groups of researchers focusing on Topics  6, 8, and 11 were also identified. 

Based on these observations, the disaster informatics field has not only attracted interest within research-focused countries but also fostered international collaborations. This collaborative trend is likely to create a norm for disaster informatics, emphasizing the importance of global cooperation in understanding, investigating, and developing solutions to mitigate the impacts of disasters on a worldwide scale.

\subsection{Theoretical Implications}

This study advances the theoretical understanding of disaster informatics in several ways. First, by analyzing a large-scale corpus of literature, it reveals how the field’s research priorities and thematic structures have shifted in the wake of the COVID-19 pandemic. The integration of bibliometric analysis and advanced transformer-based topic modeling techniques offers a more nuanced theoretical framework for understanding complex, multidisciplinary domains like disaster informatics. Furthermore, the incorporation of both disaster and crisis informatics concepts—along with public health considerations—provides a broader theoretical lens, underscoring the interdependency between these areas. By introducing the Topic Embedding Similarity metric and combining it with established evaluation measures, the study contributes a theoretical refinement of topic modeling evaluation methods, supporting more robust, theoretically informed assessments of thematic trends in large textual datasets. 

By generating human-like topic descriptions, ChatGPT provides nuanced, context-sensitive interpretations that traditional topic modeling tools alone may not fully capture. This intersection of advanced NLP and topic modeling theory brings forth a new level of semantic understanding, allowing researchers to move beyond static keyword lists toward more conceptually coherent themes. Furthermore, applying LLMs to summarize and interpret topics aligns with emerging theoretical frameworks that emphasize human-in-the-loop analytics, bridging the gap between machine-generated patterns and domain expertise. Theoretical models of knowledge discovery in text now must consider how generative AI can enhance interpretability, shape topic boundaries, and integrate broader contextual cues into the study of complex, multidisciplinary fields like disaster informatics.

% third contribution is based on the identified topics. What are the implications of the 12 topics?

\subsection{Practical Implications}

From a practical standpoint, the findings can guide policymakers, emergency managers, public health officials, and other stakeholders in developing more holistic and informed strategies for disaster preparedness, response, and recovery. The study’s methodology and results help practitioners identify emerging topics, key contributors, and leading institutions, facilitating the formation of targeted collaborations and resource allocations. Recognizing the heightened importance of public health considerations in disaster informatics enables decision-makers to design integrated approaches that address the health dimensions of crises more effectively. 

In addition, the improved topic modeling techniques and evaluation metrics presented here can be applied by researchers and organizations seeking to better understand large, complex bodies of literature, ultimately supporting evidence-based policy development and more resilient disaster management frameworks. The user-friendly, coherent summaries produced by ChatGPT make it easier for stakeholders to quickly understand and apply complex research findings, thereby facilitating data-driven decision-making in disaster preparedness, response, and recovery. 

This study streamlines the identification of key contributors, influential institutions, and geographically concentrated research efforts within specific topic areas. By enabling more targeted strategies, such as strengthening collaborative networks and focusing funding on high-impact research themes, practitioners can optimize resource allocation and enhance the effectiveness of risk reduction measures. 

In essence, integrating bibliometric and LLM's capabilities into topic modeling on large-scale academic literature analysis transforms these methods into actionable intelligence tools, delivering insights that are both theoretically sound and operationally valuable for disaster informatics and related domains.

\section{Conclusion and Future Work}
\label{sec:conclusion}
This study conducted a bibliometric analysis and content analysis of the collected disaster informatics literature from January 2020 to September 2022. The results indicate that the countries that were most impacted by the COVID-19 pandemic were also among the most active, with each country having specific research interests. The most active countries identified from the collected data were also the most active across the most significant topics. The top three countries, the U.S., India, and the UK, had strong partnerships, with each pair collaborating on a particular topic.

The most active institutions were King Saud University, the University of California, and the University of Toronto, and across the most significant topics were King Saud University, the University of Porto, and the University of Sao Paulo. In terms of collaborations, countries and institutions within the same region or share a common language tend to collaborate. Charis Galanakis was among the most active authors across the collected data and the most significant topics. In terms of partnerships, many top authors tend to form close partnerships with one or two key partners, aside from their extensive networks of occasional collaborations. Across the most significant topics, authors typically specialized in one or two specific topics, while institutions had more diverse interests across several topics. 

Topic modeling results showed that BERTopic was the most optimal model highlighting the potential for refining and adapting existing techniques. The evaluation metrics proposed in this study were effective and necessary for selecting the best topic modeling methods and topic modeling results. For topic description, Llama 2 generated better descriptions for the extracted topics, compared to GPT-3.5-turbo. 

Our content analysis results indicate an evolution of disaster informatics from a narrow focus on natural disasters with limited consideration of public health, to a more holistic approach that addresses the pandemic-related challenges across multiple fields. Overall, the COVID-19 pandemic has influenced research priorities in the field of disaster informatics, placing greater emphasis on public health.

This study has limitations. Firstly, the data was collected from January 2020 to September 2022, meaning that our results may not reflect the latest topics and trends of publications beyond September 2022. 
% the 2022 publications do not represent the entire 2022 publications. 
% Secondly, the data was collected from only three databases, excluding non-English publications. This limitation potentially restricts the comprehensiveness of our analysis within the scope of disaster informatics literature. 
% Thirdly, many publications contained formatting issues such as extra spacing, irrelevant details, and incomplete information in the affiliation fields, which complicates automatic data parsing. We addressed this by conducting annotation training, providing written instructions and feedback, and performing spot-checking. Despite these efforts, some errors in the annotation may still remain.
The quality of topic descriptions relies on the selected GenAI for summarization. Given that LLMs are known for suffering from catastrophic forgetting \citep{Luo2023AnES} and hallucination \citep{Ji2022SurveyOH}, the generated descriptions may not accurately reflect the entire content of each topic.

In the future, we will extend the data collection period after September 2022 to capture the latest trends and topics in disaster informatics. Comparative studies across different pandemics or disaster responses could provide valuable lessons in improving responses to global crises. Secondly, citation analysis can further help uncover groundbreaking studies that form the foundation of disaster informatics. We also plan to investigate collaboration patterns from an interdisciplinary perspective to understand how diverse fields converge to address complex challenges, providing insights into factors that hinder or drive such partnerships. In addition, we will develop advanced automated parsing techniques to overcome the challenges associated with manual annotations and data quality issues in publication metadata. Last but not least, we will explore newer models with improved memory and summarization capabilities, and experiment with different prompting strategies to improve the quality of topic descriptions.

\appendix
\section{Appendix}
\subsection{Detailed Description of Top 12 Most Significant Topics}
\label{sec:appendix_topic_description}
% Move to the appendix
% Below is the description of the 12 most significant topics:
\begin{itemize}
    \item Topic 1: Impact of COVID-19 on International Trade (1.98 significant score with 36 publications). It particularly focuses on trade policies \citep{Welfens_2020, Bown_2021, Vo_Le_2022}, the dynamics of import and export \citep{Bown_2021, Cerdeiro_Komaromi_2022}, the impacts on various regions and industries \citep{Welfens_2020, Marel_2021, Vo_Le_2022}, the geopolitical implications \citep{Welfens_2020, Hayakawa_Imai_2021}, and the responses of international organizations and governments during the pandemic \citep{Debre_Dijkstra_2021, Zeytoon_Nejad_Hasnain_2021}. 
    
    \item Topic 2: Burnout Among Healthcare Professionals During the COVID-19 Pandemic (1.90 significant score with 32 publications). Topic 2 discusses the well-being and mental health of professionals in healthcare and other sectors during the COVID-19 pandemic. This topic addresses critical issues such as burnout, stress, depression, and emotional exhaustion \citep{Alwashmi_Alkhamees_2021, Sklar_Ehrhart_Aarons_2021, Goncalves_Castro_Rego_Nunes_2021, Armstrong_Porter_Larkins_Mesagno_2022, Kim_Yoo_Cho_Hwang_2022}, and examines the role of personal resources, such as resilience, self-care, and self-compassion \citep{Ferreira_Gomes_2021, Sanso_Galiana_Oliver_Blanco_2020} and coping strategies, in mitigating burnout and improving mental health among healthcare workers \citep{Ferreira_Gomes_2021, Barello_Caruso_Palamenghi_2021, Canady_2022}.  

    \item Topic 3: Medical Technology and Additive Manufacturing during COVID-19 Pandemic (1.85 significant score with 37 publications). Topic 3 describes the use of technology and additive manufacturing to address the shortage of medical devices and personal protective equipment such as face masks and face shields during the COVID-19 pandemic \citep{Abbas_2021, Agarwal_2022, Belhouideg_2020, Patel_Gohil_2021}. The topic covers the development of innovative devices \citep{Munoz_Pumera_2021, Sherborne_Claeyssens_2024, Pais_Ferreira_Pires_Silva_2022}, the challenges and limitations of using these technologies in the pandemic situation \citep{Agarwal_2022, Bharti_Singh_2020, Ballardini_Mimler_Minssen_Salmi_2022, Tareq_Rahman_Hossain_Dorrington_2021}. 

    \item Topic 4: Waste Management and Environmental Impacts of COVID-19 (1.76 significant score with 37 publications). It was also recognized as the key research area in post-pandemic waste management in \citep{Ranjbari_et_al_2023}. This topic addresses the challenges and opportunities in waste management during the COVID-19 pandemic \citep{Yazdian_Jamshidi_2021, Osra_Morsy_ElRahim_2021}, focusing on various types of waste (plastic, medical, personal protective equipment, and microplastics) \citep{Haque_Fan_2022, Dehal_Vaidya_Kumar_2021b, Ribeiro_Castro_2022}, examining the environmental impacts of such waste \citep{Haque_Fan_2022, Dehal_Vaidya_Kumar_2021b, Ribeiro_Castro_2022, Du_Huang_Wang_2022}, and proposing strategies for effective waste management and mitigation of its environmental impacts \citep{Yazdian_Jamshidi_2021, Dehal_Vaidya_Kumar_2021b, Du_Huang_Wang_2022, Osra_Morsy_ElRahim_2021}.  

    \item Topic 5: Community Pharmacy Practice During COVID-19 (1.75 significant score with 31 publications). Topic 5 examines the challenges faced by community pharmacists, their preparedness, and the critical role they play in maintaining patient safety while providing essential healthcare services \citep{Gregory2020COVID19HD, Bahlol2020PandemicPO, SevillaSnchez2020PharmaceuticalCI, Kassem2021CommunityPN} during COVID-19. The topic also explores the transition to online learning for pharmacy students, highlighting adjustments in educational practices during the pandemic \citep{Hussain2021PerformanceOP, Fuller2020APS, Law2021AdaptingPE}. 

    \item Topic 6: Genomics and Variants of SARS-CoV (1.74 significant score with 38 publications). Topic 6 investigates the genetic characteristics and mutations of SARS-CoV variants \citep{Chakraborty2022ComparativeGE, Kim2020SARSCoV2EA, Ashwaq2021V483AAE}, their transmission dynamics \citep{Kim2020SARSCoV2EA, Bandoy2021AnalysisOS}, effects on disease transmission and the severity of the pandemic \citep{Jalilian2020AHS, Bandoy2021AnalysisOS, Ashwaq2021V483AAE}. 

    \item Topic 7: The Role of Nutrition in the Immune System during COVID-19 (1.67 significant score with 43 publications). Topic 7 examines how diet and nutrition influence the immune system \citep{Toor2022ExploringDA, Birk2020NutrigeneticsOA}, highlighting the benefits of specific food components \citep{Przybylska2022LycopeneIT, Barbosa2021PolysaccharidesOF, Le2020BioactiveCA} and the role of antioxidants and essential nutrients in lowering disease risk and severity \citep{Birk2020NutrigeneticsOA, Toor2022ExploringDA}.

    \item Topic 8: COVID-19 Pandemic’s Impact on Dental Education and Practice (1.64 significant score with 49 publications). Topic 8 addresses the challenges encountered by dental students \citep{Salgado2020ClinicalCS, Adam2021DentalSD} and practitioners during the pandemic \citep{Turkistani2020DentalRA, Dziedzic2020DentalCP}, and highlights the adaptations made within dental education and practice, such as shifting to online learning and introducing new infection control measures, to mitigate the effects of the pandemic \citep{Salgado2020ClinicalCS, Turkistani2020DentalRA, Dziedzic2020DentalCP}. 

    \item Topic 9: Impact of COVID-19 Pandemic on Pregnant Women and Maternal Health (1.58 significant score with 44 publications). Topic 9 covers the challenges and impacts of the pandemic on maternal health, addressing both physical and mental health aspects \citep{Dule2021PsychologicalDA, Ehsan2020AnalysingTI, Doyle2020PostnatalDR, mirzakhani2022HighriskPW}. It also examines the access to healthcare support and services by pregnant and postpartum women during this critical time \citep{Turner2022ARO, RamnMichel2022AbortionAA, mirzakhani2022HighriskPW}.

    \item Topic 10: Consumer Behavior and Trust During the COVID-19 Pandemic (1.58 significant score with 61 publications). Topic 10 explores the pandemic's impact on consumer behaviors (including panic buying, drinking at home, etc.) across various industries, such as retail, finance, tourism, and healthcare \citep{GordonWilson2021ConsumptionPD, Sheth2020ImpactOC, GordonWilson2021AnEO}. It also examines companies' responses, the resulting shifts in consumer satisfaction \citep{Baber2021EfficacyOC, Wu2022DeterminingTF}, and factors influencing consumer trust in financial institutions during the pandemic \citep{Baber2021EfficacyOC, Bijlsma2021DeterminantsOT}.  

    \item Topic 11: Virus Inactivation and Detection (1.55 significant score with 53 publications). Topic 11 investigates the detection and inactivation of bacteria and viruses using antimicrobial materials and surfaces, UV light, and nanotechnology \citep{Wainwright2021AntiinfectiveDI, Arentz2021EvaluationOM, Yasri2022SustainableMA}. It also examines the mechanisms of virus inactivation, particularly in food engineering, to improve food safety and mitigate infectious diseases like COVID-19 \citep{Roos2020WaterAP, Farahmandfar2021MonitoringON, Ceylan2021Chapter1A}. 

    \item Topic 12: The Role of Technology in the COVID-19 Pandemic (1.50 significant score with 61 publications). Topic 12 discusses the use of technology in managing the pandemic \citep{Murad2022WirelessTF}, assesses user acceptance and adaptation to these tools \citep{Troisi2021Covid19SI}, analyzes its societal impacts \citep{Lyu2022CulturalIO, Sultana2022EvaluatingTP} including privacy concerns and associated risks \citep{Sharon2020BlindsidedBP, Ioannou2021PrivacyAS}.

\end{itemize}

\subsection{Authorship in the 12 most Significant Topics}
\label{sec:appendix_authorships}

\newcolumntype{L}[1]{>{\raggedright\arraybackslash}m{#1}}
\newcolumntype{C}[1]{>{\centering\arraybackslash}m{#1}}
\renewcommand{\tabularxcolumn}[1]{>{\arraybackslash}m{#1}}

\begin{longtable}{C{0.6cm}L{5.1cm}L{5.8cm}L{2.4cm}}

\caption{Authorship in the most significant 12 Topics. * indicates that in a given topic, there are no authors (or institutions) with the highest number of publications. \dag \space indicates that there are 20 authors with only publications each. The number inside the parenthesis is the number of publications.} \label{tab:authorship_12topics} \\
\hline
\centering
\textbf{Topic ID} & \textbf{Active Authors} & \textbf{Active Institutions} & \textbf{Active Countries} \\ \hline
\endfirsthead
\caption[]{(continued)} \\
\hline
\textbf{Topic ID} & \textbf{Active Authors} & \textbf{Active Institutions} & \textbf{Active Countries} \\ \hline
\endhead
\hline \endfoot
1 & Anirudh Shingal, India (2), \newline Bernard Hoekman, Italy (2),\newline Ernesto Rodríguez-Crespo, Spain (2). & World Bank, US (2). & US (9),\newline Italy (4),\newline UK (3),\newline India (3). \\ \hline 

2 & Maria Tsironi, Greece (2), \newline Aspasia Panagiotou, Greece (2), \newline Ioannis Ilias, Greece (2), \newline Pachi Argyro, Greece (2), \newline Christos Sikaras, Greece (2), \newline Athanasios Tselebis, Greece (2), \newline Sofia Zyga, Greece (2). & University of Peloponnese, Greece (2), \newline Elena Venizelou Hospital, Greece (2). & US (3), \newline South Korea (3),\newline Switzerland (3), \newline Greece (3), \newline Italy (3). \\ \hline 

3 & *(1) & *(1) & US (6), \newline India (5), \newline UK (5). \\ \hline 

4 & Charis Galanakis, Greece (3), \newline Turki Aldawoud, Saudi Arabia (2), \newline Kit Wayne Chew, Singapore (2). & King Saud University, Saudi Arabia (3). & India (7), \newline China (6), \newline Malaysia (4). \\ \hline 

5 & Mohamed Bahlol, Egypt (2), \newline Rebecca Susan Dewey, UK (2), \newline  Alamdar Hussain, US (2). & University of Toronto, Canada (2), \newline Egyptian-Russian University, Egypt (2), \newline University of South Carolina, US (2), \newline American University of Health Sciences, US (2), \newline King Abdulaziz University, Saudi Arabia (2), \newline University of Nottingham, UK(2). & US (8), \newline UK (5), \newline Saudi Arabia (3), \newline Portugal (3), \newline Australia (3). \\ \hline 

6 & Manojit Bhattacharya, India (3), \newline Chiranjib Chakraborty, India (3), \newline Sang Soo Lee, US (3), \newline Ashish Ranjan Sharma, South Korea (3). & Hallym University-Chuncheon Sacred Heart Hospital, South Korea (3), \newline Fakir Mohan University, India (3), \newline Adamas University, India (3). & India (7), \newline South Korea (6), \newline China (6), \newline US (5), \newline Saudi Arabia (4). \\ \hline 

7 & Charis Galanakis, Greece (3), \newline Salam Ibrahim, Saudi Arabia (2). & North Carolina Agricultural and Technical State University, US (4), \newline Galanakis Laboratories, Greece (3), \newline King Saud University, Saudi Arabia (2). & India (12), \newline US (8), \newline Saudi Arabia (4), \newline Poland (4), \newline Greece (4). \\ \hline 

8 & Arkadiusz Dziedzic, Poland (3), \newline Marta Tanasiewicz (2). & Medical University of Silesia, Poland (3), \newline University of Pennsylvania, US (2), \newline Datta Meghe Institute of Medical Sciences, India (2), \newline King Saud University, Saudi Arabia (2). & US (12), \newline Saudi Arabia (9), \newline India (8). \\ \hline 

9 & Darmawati, Indonesia (2), \newline Aman Dule, Ethiopia (2) & Mettu University, Ethiopia (2), \newline Syiah Kuala University, Indonesia (2), \newline University of Porto, Portugal (2), \newline Jordan University of Science and Technology, Jordan (2). & US (7), \newline Italy (5), \newline UK (4), \newline Ethiopia (4), \newline Portugal (4), \newline Iran (4). \\ \hline 

10 & Angelika Ploeger, Germany (3), \newline Xin Qi, Germany (3). & University of Portsmouth, UK (3), \newline University of Kassel, Germany (3). & US (8), \newline UK (8), \newline China (8), \newline Italy (6), \newline Malaysia (4), \newline South Korea (4). \\ \hline 

11 & \dag (2) & Port Said University, Egypt (3), \newline Duksung Women's University, South Korea (2), \newline Liverpool John Moores University, UK (2), \newline Technical University of Braunschweig, Germany (2), \newline University of Campania Luigi Vanvitelli, Italy (2), \newline University of Coimbra, Portugal (2), \newline University in Nizwa, Oman (2), \newline Zagazig University, Egypt (2). & India (11), \newline US (6), \newline UK (5), \newline Russia (4), \newline Brazil (4). \\ \hline 

12 & Lutz Peschke, Turkey (3), \newline Frans Folkword, Spain (2), \newline Francisco Lupiáñez-Villanueva (2), \newline Nahida Sultana, Bangladesh (2), \newline Marzia Tamanna, Bangladesh (2). 
& Open University of Catalonia, Spain (2), \newline Indian Institute of Technology, India (2), \newline Bangladesh University, Bangladesh (2), \newline East West University, Bangladesh (2), \newline University of the Pubjab, Pakistan (2), \newline Tilburg Univeristy, Netherlands (2), \newline Bilkent University, Turkey (2). 
& US (11),\newline Spain (7), \newline UK (7), \newline China (6), \newline India (6).\\ \hline
\end{longtable}

\printcredits

\section*{Declaration of generative AI and AI-assisted technologies in the writing process}

During the preparation of this work the authors used ChatGPT o1 in order to improve the readability and language of the manuscript. After using this tool/service, the authors reviewed and edited the content as needed and takes full responsibility for the content of the published article.

\begin{comment}
Appendix sections are coded under \verb+\appendix+.

\verb+\printcredits+ command is used after appendix sections to list 
author credit taxonomy contribution roles tagged using \verb+\credit+ 
in frontmatter.

\printcredits
\end{comment}

%\bibliographystyle{cas-model2-names}

\bibliography{cas-refs}


\begin{comment}
    \bio{draftpic1}
Author biography without author photo.
Author biography. Author biography. Author biography.
Author biography. Author biography. Author biography.
Author biography. Author biography. Author biography.
Author biography. Author biography. Author biography.
Author biography. Author biography. Author biography.
Author biography. Author biography. Author biography.
Author biography. Author biography. Author biography.
Author biography. Author biography. Author biography.
Author biography. Author biography. Author biography.
\endbio

\bio{draftpic1}
Author biography with author photo.
Author biography. Author biography. Author biography.
Author biography. Author biography. Author biography.
Author biography. Author biography. Author biography.
Author biography. Author biography. Author biography.
Author biography. Author biography. Author biography.
Author biography. Author biography. Author biography.
Author biography. Author biography. Author biography.
Author biography. Author biography. Author biography.
Author biography. Author biography. Author biography.
\endbio

\bio{draftpic1}
Author biography with author photo.
Author biography. Author biography. Author biography.
Author biography. Author biography. Author biography.
Author biography. Author biography. Author biography.
Author biography. Author biography. Author biography.
\endbio
\end{comment}

\end{document}